\documentclass[a4paper,11pt]{article}
\pdfoutput=1
\usepackage{jheppub} 
\usepackage[T1]{fontenc} 

\usepackage{verbatim}
\usepackage{enumerate}
\usepackage{mathrsfs}
\usepackage{bm}
\usepackage{tikz}
\usepackage{pgfplots}
\usepackage{subfig}

\newcommand{\tr}{\text{tr}}

\newcommand{\Del}{\nabla}
\newcommand{\del}{\partial}
\newcommand{\dd}{\text{d}}

\renewcommand{\tilde}{\widetilde}

\usetikzlibrary{arrows, positioning, shapes.geometric, decorations.markings, patterns}
\usepackage{tikz-3dplot}

\newcommand{\bra}[1]{\langle #1 |}
\newcommand{\ket}[1]{| #1 \rangle}

\def\fig#1{Figure \ref{#1}}

\definecolor{VHcol}{rgb}{0.7,0.3,0.9}

\def\regA{{A}}  
\def\regB{{B}}  
\def\regAB{{\regA \cup \regB}}  
\def\rhoA{\rho_{\regA}}  
\def\extr#1{\mathcal{E}_{#1}}  
\def\extrA{\mathcal{E}_{\regA}}  
\def\EWA{\mathcal{W}_{\regA}}  
\def\HR#1{\mathcal{R}_{#1}}  
\def\HRA{\mathcal{R}_{\regA}}  
\def\IAB{I(\regA:\regB)}  
\def\volAB{{\mathcal V_{\regAB}}} 

\def\ZZ{{\mathbb{Z}}} 
\def\sN{{\sf{N}}}  

\title{Large-$d$ phase transitions in holographic mutual information}

\author[1]{Sean Colin-Ellerin,}
\author[1]{Veronika E. Hubeny,}
\author[2]{Benjamin E. Niehoff,}
\author[3]{and Jonathan Sorce}

\affiliation[1]{Center for Quantum Mathematics and Physics (QMAP), Department of Physics, University of California, Davis, CA 95616, U.S.A.}
\affiliation[2]{Instituut voor Theoretische Fysica, KU Leuven, Celestijnenlaan 200D, B-3001 Leuven, Belgium}
\affiliation[3]{Stanford Institute for Theoretical Physics, Stanford University, 382 Via Pueblo Mall, Stanford, CA 94305-4060, U.S.A.}

\emailAdd{scolinellerin@ucdavis.edu}
\emailAdd{veronika@physics.ucdavis.edu}
\emailAdd{ben.niehoff@kuleuven.be}
\emailAdd{jsorce@stanford.edu}

\abstract{In the AdS/CFT correspondence, the entanglement entropy of subregions in the boundary CFT is conjectured to be dual to the area of a bulk extremal surface at leading order in $G_N$ in the holographic limit. Under this dictionary, distantly separated regions in the CFT vacuum state have zero mutual information at leading order, and only attain nonzero mutual information at this order when they lie close enough to develop significant classical and quantum correlations. Previously, the separation at which this phase transition occurs for equal-size ball-shaped regions centered at antipodal points on the boundary was known analytically only in $3$ spacetime dimensions. Inspired by recent explorations of general relativity at large-$d$, we compute the separation at which the phase transition occurs analytically in the limit of infinitely many spacetime dimensions, and find that distant regions cannot develop large correlations without collectively occupying the entire volume of the boundary theory. We interpret this result as illustrating the spatial decoupling of holographic correlations in the large-$d$ limit, and provide intuition for this phenomenon using results from quantum information theory. We also compute the phase transition separation numerically for a range of bulk spacetime dimensions from $4$ to $21$, where analytic results are intractable but numerical results provide insight into the dimension-dependence of holographic correlations. For bulk dimensions above $5$, our exact numerical results are well approximated analytically by working to next-to-leading order in the large-$d$ expansion.}

\makeatletter
\def\@fpheader{\vspace{0.2cm}}
\makeatother

\begin{document} 
\maketitle
\flushbottom

\section{Introduction and Summary}
\label{sec:intro}

Entanglement has played an increasingly prominent role in theoretical physics,  particularly in the context of holographic dualities.
The broad expectation that the emergence of a dynamical bulk spacetime will be elucidated by a better understanding of the structure of entanglement in the boundary field theory has motivated many vibrant explorations into holographic entanglement, often utilizing the bulk geometry.
One useful measure of entanglement in a pure state is the \emph{entanglement entropy}, defined as the von Neumann entropy of the reduced density matrix associated with a bipartition of the full Hilbert space.\footnote{\ 
Although the quantum field theory Hilbert space does not strictly-speaking factorize, this will not impact the present discussion; for a nice summary of the associated subtleties see e.g.\ \cite{Witten:2018lha}.
}
In a local quantum field theory, it is natural to partition the system into spatial regions; for a theory which admits a holographic dual, in a state describing a classical bulk geometry,  entanglement entropy can be obtained using the Ryu-Takayanagi (RT) \cite{RT} (and its covariant generalization, the Hubeny-Rangamani-Takayanagi (HRT) \cite{HRT}) prescription (often collectively referred to as HRRT), which expresses this quantity geometrically, in terms of an area of a bulk surface. More specifically, the entanglement entropy $S(\regA)$ of a given boundary region $\regA$ is given to leading order in a small-$G_N$ expansion by the quarter-area of the smallest-area extremal surface $\extrA$ homologous to that region: 
\begin{equation}\label{eq:HRT}
S(\regA) = \min_{\extrA \sim \regA} \left[ \frac{1}{4 G_N}\, \text{Area}(\extrA) \right] \ + O(1).
\end{equation}	
In concrete realizations of the AdS/CFT correspondence, this expression has been argued to follow from path integral arguments \cite{LM,DLR}, with the $O(1)$ corrections having been computed in \cite{FLM}. The geometrization of entanglement given by equation \eqref{eq:HRT} has led to many intriguing insights and developments; see e.g.\ \cite{Rangamani:2016dms} for a recent review.

Although the entanglement entropy of a spatial region with nonempty boundary suffers from a UV divergence in any local quantum field theory (its bulk dual manifested by the associated HRRT surfaces having infinite proper area since they reach the spacetime boundary), one can nevertheless construct meaningful finite quantities by combining entanglement entropies of several subsystems so that the UV divergences cancel in a cutoff-independent way \cite{cutoffs}. The most basic such quantity is the mutual information $\IAB$, defined by
\begin{equation}\label{eq:MIdef}
 \IAB = S(\regA) + S(\regB) - S(\regAB) \ .
\end{equation}	
The mutual information measures how the individual entropies of regions $\regA$ and $\regB$ differ from their joint entropy. In this sense, it characterizes the total amount of correlation between the subsystems $\regA$ and $\regB$. By using the HRRT formula \eqref{eq:HRT} to compute the mutual information for various subsystems in a holographic state, one can elucidate how the state is `held together' by its spatial correlations.

\begin{figure}[h]
\centering
\subfloat[\label{fig:AdS3-disconnected}]{
\begin{tikzpicture}
\centering
\filldraw[fill=gray!10!white, draw=black!70!white] (-4.5,0) circle (3cm);
\draw[blue,thick] (-2.38,2.12) arc (45:135:3cm);
\draw[blue,thick] (-6.62,-2.12) arc (225:315:3cm);
\node at (-4.5,3.5) {\scalebox{1.1}{$\color{blue}\regA$}};
\node at (-4.5,-3.5) {\scalebox{1.1}{$\color{blue}\regB$}};
\fill[fill=green!25!, opacity = 0.8] (-6.61,2.12) .. controls (-5.4,0.95) and (-3.6,0.95) .. (-2.38,2.12) .. controls (-3.6,3.28) and (-5.4,3.28) .. (-6.62,2.12)--cycle;
\fill[fill=green!25!, opacity = 0.8] (-6.61,-2.12) .. controls (-5.4,-0.95) and (-3.6,-0.95) .. (-2.38,-2.12) .. controls (-3.6,-3.28) and (-5.4,-3.28) .. (-6.62,-2.12)--cycle;
\draw[black] (-6.62,2.12) arc (225:315:3cm);
\draw[black] (-2.38,-2.12) arc (45:135:3cm);
\node at (-4.5,0.8) {\scalebox{1.1}{$\mathcal{E}_{\regA}$}};
\node at (-4.5,-0.8) {\scalebox{1.1}{$\mathcal{E}_{\regB}$}};
\end{tikzpicture}
}
\hspace{2cm}
\subfloat[\label{fig:AdS3-connected}]{
\begin{tikzpicture}
\filldraw[fill=gray!10!white, draw=black!70!white] (4.5,0) circle (3cm);
\draw[blue,thick] (6.62,2.12) arc (45:135:3cm);
\draw[blue,thick] (2.38,-2.12) arc (225:315:3cm);
\node at (4.5,3.5) {\scalebox{1.1}{$\color{blue}\regA$}};
\node at (4.5,-3.5) {\scalebox{1.1}{$\color{blue}\regB$}};
\fill[fill=magenta!25!, opacity = 0.8] (2.38,2.12) .. controls (3.55,1.05) and (3.55,-1.05) .. (2.38,-2.12) .. controls (3.6,-3.28) and (5.4,-3.28) .. (6.62,-2.12) ..controls (5.45,-1.05) and (5.45,1.05) .. (6.62,2.12) .. controls (5.4,3.28) and (3.6,3.28) .. (2.38,2.12)--cycle;
\draw[black] (6.62,2.12) arc (135:225:3cm);
\draw[black] (2.38,-2.12) arc (315:405:3cm);
\node at (4.5,1) {\scalebox{1.1}{$\extr{\regAB}$}};
\draw[->] (4.5,0.6) -- (3.4,0.2);
\draw[->] (4.5,0.6) -- (5.6,0.2);
\end{tikzpicture}
}
\caption{Phase transition in the HRT surface and corresponding entanglement wedge (whose spatial slice is indicated by the shaded homology region $\HR{\regAB}$) for the boundary region $\regAB$ in AdS$_{3}$. (a) Shows the decorrelated phase, while (b) shows the correlated phase. When $\regA$ and $\regB$ each take up one quarter of the boundary length, the surfaces $\extr{\regAB}$ and $\extr{\regA} \cup \extr{\regB}$ have equal area.}
\label{f:AdS3}
\end{figure}
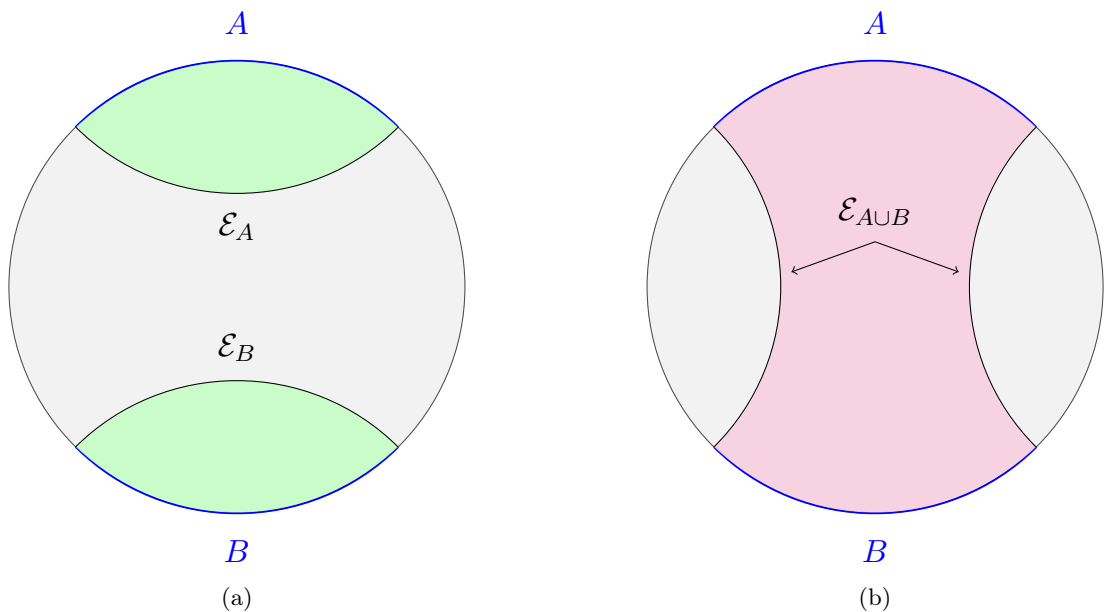

Due to a universal relation known as subadditivity, the mutual information cannot be negative; i.e., $\IAB \ge 0$ holds for any subpartition of any physically allowed state.  Indeed, in an ordinary quantum field theory living on a single connected background, ${\IAB}$ is strictly positive: the mutual information of two regions is bounded below by connected correlation functions of operators supported on the two regions \cite{Wolf:2007tdq,Cardy:2013nua}, and these correlation functions are generally nonvanishing. That said, the fact that the HRRT formula \eqref{eq:HRT} is expressed in terms of an expansion in $G_N$ indicates a \emph{hierarchy of correlation scales} in any holographic theory. Subregions of a geometric state in a holographic field theory can be correlated only at $O(1)$, with vanishing correlations at order $G_N^{-1}.$ In fact, the minimality condition in equation \eqref{eq:HRT} makes it easy to engineer subregions that satisfy $\IAB = 0$ at leading order in $G_N$. As the separation between any two spatial regions is increased, keeping the state fixed, the HRT surface $\extr{\regAB}$ computing entanglement entropy of the joint system $\regAB$ undergoes a phase transition from a connected surface (joining the boundaries of $\regA$ and $\regB$) to a pair of disconnected surfaces $\extrA \cup \extr{\regB}$. These two configurations are sketched in Figure \ref{f:AdS3} for antipodal boundary regions in a time slice of vacuum AdS$_{3}$. The relation $\IAB=0$ at order $G_N^{-1}$ thus follows directly from equation \eqref{eq:HRT}.\footnote{\
This characteristic property of geometric states in holography was recently utilized by \cite{Hubeny:2018ijt,Hubeny:2018trv} in constructing the multi-party entanglement relations delineating the holographic entropy cone (initially explored in \cite{Bao:2015bfa}).}

The hierarchy of correlation scales in small-$G_N$ holographic systems, together with the presence of this ``phase transition'' in which separated regions may spontaneously develop correlations at order $G_N^{-1}$ upon tuning their separation, indicates a rich structure of coupling (and decoupling) in the spatial correlations of holographic states. This phenomenon of spatial decoupling is the central focus of this paper. We investigate the rich correlation structure of holographic theories by analyzing the mutual information phase transition in more than $3$ bulk dimensions, which has until now been the only analytically tractable case.\footnote{\ 
While certain aspects of connected and disconnected extremal surfaces were investigated in $4$ bulk dimensions in \cite{krtous}, the phase transition itself was not the object of study.} In particular, we will find an analytic solution for the phase transition in the limit $d \to \infty$. One may object that the AdS$_{d+1}$/CFT$_d$ correspondence is microscopically well-defined only for a rather limited set of $d$'s (and in fact, interacting CFTs are only known to exist for $d \le 6$). However, the bulk geometry is of course well-defined at any $d$, and some effective quantum system (such as a tensor network) may well still utilize the `holographic' crutch of computing entanglement entropy by bulk extremal surface areas.\footnote{\ 
Moreover, since the HRRT formula can be viewed as a generalization of the 
black hole Bekenstein-Hawking entropy formula to arbitrary spatial partitions of the boundary, one might well expect it to hold for general holographic theories, not just those that admit a microscopic realization in terms of a strongly interacting CFT.}  We will therefore {\it assume} that equation \eqref{eq:HRT} is meaningful at all $d$, and in particular that it allows us to relate the presence of correlations between spatially separated regions in the boundary with the connectivity of HRT surfaces in the bulk.

Our analysis is motivated partially by recent developments in the theory of large-$d$ general relativity \cite{largeD}, which has proved a surprisingly powerful computational (as well as conceptual tool) for approximating various quantities in the more physically relevant context of $d=4$ or $5$ in terms of an expansion in $1/d$. The essential insight of the large-$d$ program in general relativity is that at infinite $d$, the complicated non-linear physics of Einstein's equations simplifies considerably. This can be seen heuristically in the fact that the Newtonian potential in $(d+1)$ dimensions scales as $1/r^{d-2}$; massive objects, even ones that are quite near one another, tend to \emph{decouple} from one another at large $d$. Our explorations below will show, in precise terms, that the holographic mutual information obeys a similar principle: spatial regions tend to decouple from one another in the large-$d$ limit.

The mutual information phase transition is of interest not only because it illustrates the phenomenon of spatial decoupling, but because it underlies another interesting feature of holography related to bulk reconstruction.  It is believed that the bulk dual of the reduced density matrix $\rhoA$ associated to a given spatial region $\regA$ on the boundary is its entanglement wedge\footnote{\ 
The term ``entanglement wedge'' was coined and defined in \cite{Headrick:2014cta}, where the authors used it to prove consistency of HRT with boundary causality and proposed it as the natural dual of $\rhoA$.  Closely related constructs were previously considered by \cite{maximin} and \cite{Czech:2012bh}; the latter work having originally specified `the gravity dual of a density matrix' as the bulk region whose geometry remains invariant under a restricted variation of the dual CFT state $\rho$ which keeps the reduced density matrix $\rhoA$ fixed.
Subsequent support for the conjecture that the dual of $\rhoA$ is  $\EWA$ was provided in \cite{DHW,noisyDHW}.
} 
$\EWA$, namely the bulk domain of dependence of the `homology region'  $\HRA$ which itself is defined as a spatial region whose only boundaries are $\regA$ and its bulk HRT surface $\extrA$.  Since the HRT surface of a region can `jump' upon tuning its size, the corresponding entanglement wedge can correspondingly jump under continuous variations of $\regA$ (as well as under variations of the state). Such a jump naturally occurs in the presence of a phase transition in the order $G_N^{-1}$ mutual information between two regions.

As a simple example, consider two intervals $\regA$ and $\regB$ positioned antipodally on a $t=0$ slice of pure global AdS$_{3}$, as sketched in \fig{f:AdS3}. The homology requirement in \eqref{eq:HRT} allows two possible configurations for the HRT surface for the composite subsystem $\regAB$.  One (Fig.\ref{fig:AdS3-disconnected}) with a disconnected homology region  bounded by $\extrA$ and $\extr{\regB}$, and one (Fig.\ref{fig:AdS3-connected}) with a connected homology region bounded by $\extr{\regAB}$. When the two intervals jointly cover half of the boundary circle, the two sets of HRT surfaces have the same area (by symmetry), so this configuration lies precisely on the phase transition.  Increasing the size of $\regAB$ (or bringing its components closer together by breaking the $\ZZ_4$ symmetry) pushes the configuration into the connected (and hence correlated) phase with $\IAB = O(G_N^{-1})$, while conversely shrinking the size of $\regAB$ pushes the configuration into the disconnected (decorrelated) phase with $\IAB =O(1)$.  

That the entanglement wedge may jump upon a small change in the parameters of $\regAB$ implies that the density matrix $\rho_\regAB$ has a rather intricate structure, encoding the bulk geometry in a highly non-trivial way.  Indeed, the size  of the bulk region $\EWA$ which is associated to a given boundary subsystem $\rhoA$ (with $\regA$ now specifying an arbitrary collection of regions) can jump by an arbitrarily large amount under continuous variations of $\regA$.  One way to see this is to consider partitioning the boundary circle into $2 \, \sN$ equal intervals and taking $\regA$ to consist of every other interval, i.e.\  the $\sN$ non-adjoining intervals which collectively cover half the boundary.  Then, analogously to the situation sketched in \fig{f:AdS3}, in the $\sN \to \infty$ limit, the corresponding homology region would jump, upon changing the size of a single interval, from covering `almost everything' to  `almost nothing' on the Poincar\'{e} disk.\footnote{\ 
In the global AdS conformal diagram, the corresponding entanglement wedge $\EWA$ would jump from the Wheeler-de Witt patch of the given time slice to `almost nothing' consisting of the $\sN$ individual single-interval entanglement wedges.  (Of course the UV divergent pieces would still match between the two possibilities, but the finite difference in spacetime volume would grow arbitrarily large with $\sN$.)
}  While this suggests that bulk reconstruction right {\it at} the phase transition is under qualitatively less control than otherwise, in a generic situation (namely away from the co-dimension $\ge$1 set of fine-tuned configurations where several families of extremal surfaces have exactly the same areas), this subtlety does not arise.  Nevertheless, the {\it presence} of phase transitions remains even if we restrict attention to generic configurations for purposes of bulk reconstruction, so it is of interest to identify {\it where} in the parameter space such phase transitions occur. Identifying these phase transitions in parameter space, both for intermediate dimensions and in the large-$d$ limit, is the focus of the present paper.

In investigating the mutual information phase transition, with an eye toward understanding both bulk reconstruction and the decoupling of holographic correlations, we will restrict our attention to the vacuum state of a $d$-dimensional holographic field theory (and the corresponding spacetime, vacuum AdS$_{d+1}$). This state is not only analytically approachable using geometric methods, it also constitutes a ``worst-case scenario'' for investigating decoupling; the ground state is in some sense the hardest-to-decorrelate state in any given field theory. For example, as we increase the temperature in the family of thermal states of a holographic CFT, corresponding to increasing the mass of Schwarzschild-AdS black holes in the gravitational dual, the mutual information between any two fixed regions decreases.  This can be checked explicitly in the 3-dimensional case of BTZ, but it can also be easily motivated in general by the observation that at large temperature, the entanglement entropy has a volume-law scaling with region size.\footnote{\ 
In the bulk, the HRT surfaces hug the horizon, which in the conformal diagram is very close to the boundary, so that the difference between $\extr{\regAB}$ and the union of $\extr{\regA}$ and $\extr{\regB}$ gives only an (exponentially) small contribution to $\IAB$.
}

Our goal, then, is to determine the behavior of the mutual information phase transition in the ground states of holographic CFTs. As a geometrical problem, we aim to determine the separation at which two antipodal, equal-size, ball-shaped regions on the boundary of vacuum AdS$_{d+1}$ develop nonzero mutual information at order $G_N^{-1}$. This choice of boundary regions maximizes the minimal separation between the two regions while also maximizing their volume. Since the mutual information increases as the separation between the regions decreases, antipodally placed balls are the natural configuration to consider when investigating spatial decoupling, since these regions will maximize the total fraction of the boundary volume that can be occupied by $\regAB$ before developing order $G_N^{-1}$ correlations between $\regA$ and $\regB$. Furthermore, since any two ball-shaped regions can be mapped to non-antipodal regions by a conformal transformation, under which the mutual information is invariant, knowledge of the parameters of the phase transition for equal-size, antipodal balls can be used to compute the parameters of the phase transition for arbitrary pairs of ball-shaped regions on the boundary (cf.\ Appendix \ref{app:cross-ratio}).

The question, then, is as follows: how close do two antipodal, equal-size, ball-shaped boundary regions have to be in order to develop correlations at order $G_N^{-1}$? Since the geometry is static, we can use the RT prescription and limit consideration to a single time slice in the bulk, wherein we consider (spatial codimension one) minimal surfaces boundary-anchored on the pair of entangling surfaces $\partial \regA$ and $\partial \regB$.  If the angular separation between the caps is small enough, the requisite minimal surfaces will be the disconnected ones $\extr{\regA}$ and $\extr{\regB}$, whereas in the other regime there will be a single connected `tube-like' surface $\extr{\regAB}$ straddling  $\partial \regA$ and $\partial \regB$.  The location of the phase transition is the angle at which the two sets of minimal surfaces have the same area.  In the AdS$_{3}$ case of \fig{f:AdS3}, the phase transition occurs when the angular separation between the surfaces is $\pi/2$; equivalently, the phase transition occurs when the two caps collectively occupy $1/2$ of the boundary volume. In the remainder of this paper, we compute the angular separation at which antipodal regions develop order $G_N^{-1}$ mutual information (i) numerically for a range of dimensions $d > 2$ and (ii) analytically in the limit $d \rightarrow \infty.$

With this stated as our goal, let us turn to considering how the mutual information \emph{should} behave under varying $d$. We present two lines of argument, each of which would lead us to a different conclusion. First, given a system of finite total volume (which we can normalize to $\mathcal{V}=1$) in its ground state, one might naively expect that the volume $\volAB$ spanned by the subsystem $\regAB$ such that $\IAB=O(1)$ must be capped at $\volAB \le 1/2$, or some other $O(1)$ portion of the boundary. This reasoning follows from a heuristic understanding of the \emph{monogamy of entanglement}: if the degrees of freedom in $\regA$ and $\regB$ are individually in highly entangled states, without being significantly entangled to one another, then they must be entangled with distinct systems in the complement ${\overline{\regAB}}$. To support the large amount of entanglement in $\regA$ and $\regB$, the complement ${\overline{\regAB}}$ must have at least as many degrees of freedom as $\regAB$ --- or, allowing for the states on $\regA$ and $\regB$ to be submaximally entangled, perhaps some order-one fraction thereof.\footnote{\ 
This line of reasoning was suggested to us by Geoff Penington.}

Another line of argument following from monogamy of entanglement, however, suggests a very different result: if the local degrees of freedom in the holographic theory are entangled in isotropic fashion, then the higher the dimensionality, the smaller the amount of entanglement which can be spared for any given direction (since it has to be equally distributed amongst all the directions).  In the present setup, there is only a single direction (the angle) spanning the separation between $\regA$ and $\regB$; the entanglement in this direction should be greatly suppressed relative to the $d-2$ directions along the entangling surfaces  $\partial \regA$ and $\partial \regB$.  So in the large-$d$ limit, this line of argument would suggest that regions $\regA$ and $\regB$ should disentangle from each other even when they are very close (parametrically in $1/d$). This is, in fact, the line of reasoning that will be supported by our geometric arguments: we find, in Section \ref{sec:geometry}, that the angular separation at which antipodal caps develop order $G_N^{-1}$ correlations scales as $1/d$; this is in contrast to the slower, $1/\sqrt{d}$ scaling that would signal the development of large correlations at some finite fraction of the boundary volume.\footnote{\
Cf. footnote \ref{fn:sqrt-d-scaling} for the calculation that establishes this scaling.} This means that in fact $\volAB \to 1$ as $d \to \infty$ --- in other words, at large dimension, the regions decorrelate from each other even though they subtend almost the entire system!

The arguments presented thus far are largely heuristic. In the remainder of this paper, we make them concrete. In Section \ref{sec:geometry}, we use \eqref{eq:HRT} to reduce the problem of determining the mutual information phase transition to the geometric problem of comparing the areas of two families of surfaces, which in turn we reduce to the problem of solving a first-order, inhomogeneous differential equation. We then solve this equation analytically in the limit $d \rightarrow \infty$ to demonstrate the decoupling of holographic correlations at large $d$, solve it numerically at a range of smaller dimensions to show the phenomenon of decoupling at finite $d$, and finally cast all of our results in terms of conformally invariant quantities. In Section \ref{sec:monogamy}, we return to the heuristic ``monogamy of entanglement'' argument of the previous paragraph and make it precise; using theorems from the quantum information literature, we argue that isotropic systems should generically experience a spatial decoupling of entanglement at large $d$, and apply this observation to understanding the holographic results of Section \ref{sec:geometry}. In Section \ref{sec:discussion}, we interpret our results and suggest possible directions for future work.

\section{Holographic Mutual Information in the Vacuum}
\label{sec:geometry}

In this section, we derive the parameters of the mutual information phase transition in vacuum AdS$_{d+1}$ (i) analytically in the limit of large $d$, and (ii) numerically for intermediate dimensions $3 \le d \le 20$. Concretely, we will consider antipodal, equal-size, ball-shaped regions in a time slice of AdS$_{d+1}$, and find the size at which the mutual information between the two regions becomes nonzero at order $G_N^{-1}$ using the HRRT formula \eqref{eq:HRT}. Our first step will be to introduce a system of coordinates in which the differential equation satisfied by extremal surfaces homologous to these regions takes a simple form. We then study this equation analytically in a large-$d$ expansion to find the quantitative behavior of the phase transition in the large-$d$ limit, and give numerical results for dimensions $d=3$ through $d=20$ where analytical results are intractable. Finally, since the mutual information between boundary subregions in a conformal field theory is invariant under a conformal transformation of the boundary, we cast our results in terms of a conformally invariant quantity that measures the separation between the ball-shaped regions; this quantity can be used to compute the mutual information between any two ball-shaped regions by computing the mutual information of antipodal regions with the same conformal invariant. We find that in the large-$d$ limit, the boundary separation at which antipodal regions develop nonzero mutual information at order $G_N^{-1}$ vanishes as $1/(d-2)$ in the usual spherical metric for the boundary.

\subsection{Choice of Coordinates}
\label{sec:coordinates}

Consider anti-de Sitter spacetime in $d+1$ dimensions with global time coordinate $t$. Since vacuum AdS spacetime is static --- i.e., symmetric under both time translation and time reversal --- the extremal surface that computes the entanglement entropy of any region in a constant-$t$ slice of the boundary must lie entirely within the corresponding constant-$t$ slice of the bulk. In using the HRRT formula \eqref{eq:HRT} to compute the mutual information between regions in a single time slice of the vacuum state, we may therefore restrict our consideration to the constant-$t$, spacelike bulk geometry given by hyperbolic space in $d$ dimensions ($\mathbb{H}^d$).

In units where the radius of curvature is equal to one, the metric of $\mathbb{H}^d$ may be written in the Poincar\'{e} ball model as
\begin{equation} \label{eq:poincare-ball}
	ds^2 = \frac{4}{(1 - r^2)^2} \left( dr^2 + r^2 d \Omega^2_{d-1} \right),
\end{equation}
where $r$ is a radial coordinate in the range $r \in [0, 1)$ and $d \Omega^2_{d-1}$ is the metric of the $(d-1)$-sphere. The Poincar\'{e} ball model is useful in two dimensions because geodesics of the metric are given by segments of circles in the ambient space that intersect the boundary of the Poincar\'{e} ball, $r=1$, orthogonally. (By ``ambient space,'' we mean the plane generated by allowing $r$ to range from $0$ to $\infty$.) In dimensions $d > 2$, we may generalize the notion of a geodesic to that of a codimension-$1$ extremal surface. While there are many different classes of codimension-$1$ extremal surfaces in $\mathbb{H}^d$, it remains true that any codimension-$1$ surface that can be extended to form a $(d-1)$-sphere in the ambient space that intersects $r=1$ orthogonally is locally extremal. We call such a surface a \emph{minimal cap}.

\begin{figure}[h]
\centering
\subfloat[\label{fig:disconnected-min-surface}]{
%
%
%
%
%
\includegraphics{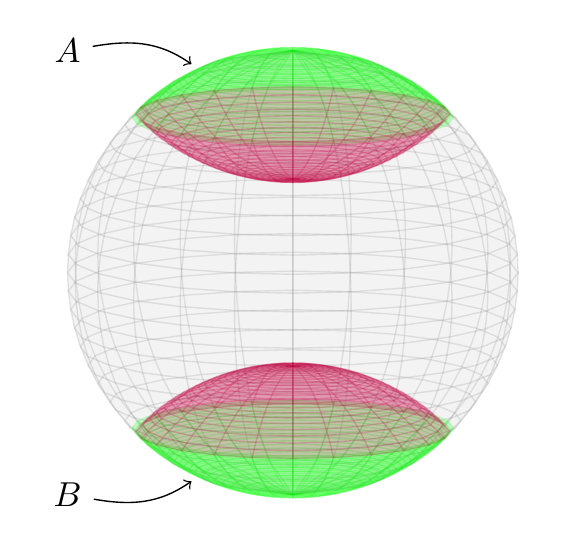}
}
\hspace{0.5cm}
\subfloat[\label{fig:connected-min-surface}]{
%
%
%
%
\includegraphics{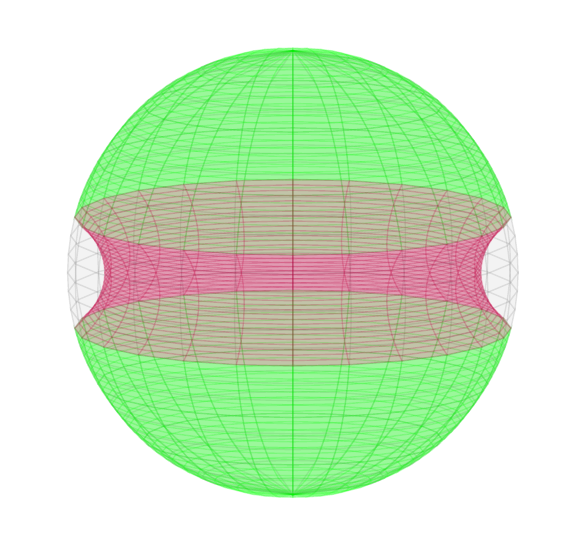}
}
	\caption{Antipodal, ball-shaped regions of equal size on the boundary of $\mathbb{H}^3.$ (a) When $A$ and $B$ are small, the minimal surface homologous to $A \cup B$ consists of two disjoint minimal caps homologous to $A$ and $B$ individually. (b) When $A$ and $B$ are large, the minimal surface homologous to $A \cup B$ is a connected, ``tube-like'' surface.}
	\label{fig:antipodal-regions}
\end{figure}

Our goal is to find the phase transition between vanishing and nonzero mutual information at order $G_N^{-1}$ for equal-size, antipodal, ball-shaped regions on the boundary. Let $\theta$ be an arbitrary choice of angular direction in the $(d-1)$-sphere, and let $A$ and $B$ be two such boundary regions centered around the points $\theta=0$ and $\theta=\pi$, respectively. These regions are sketched for $d=3$, in which case the boundary is topologically $S^2$, in Figure \ref{fig:antipodal-regions}. The minimal surfaces homologous to $A$ and $B$ are simply minimal caps in $\mathbb{H}^d$ (which, again, is itself a time-slice of AdS$_{d+1}$). Finding the minimal surface homologous to $A \cup B$, however, is more subtle. When $A$ and $B$ are sufficiently small, the minimal surface homologous to $A \cup B$ is simply the union of the disjoint minimal surfaces homologous to $A$ and $B$ individually. In this case, the mutual information
\begin{equation} \label{eq:MI}
	I(A : B) = S(A) + S(B) - S(A \cup B)
\end{equation}
vanishes at order $G_N^{-1}$ under the HRRT formula \eqref{eq:HRT}. As the regions $A$ and $B$ are made larger, however, there exist other locally extremal surfaces homologous to $A \cup B$ that could potentially be the global minima used to compute the entanglement entropy --- namely the connected, ``tube-shaped'' surfaces sketched in Figure \ref{fig:connected-min-surface}. The mutual information phase transition happens when $A$ and $B$ are large enough that one of these connected surfaces replaces the disconnected, ``two caps'' surface as the global minimum homologous to $A \cup B$. In this case, the mutual information \eqref{eq:MI} becomes positive at leading order in the HRRT formula \eqref{eq:HRT}.

Finding the precise size at which $A$ and $B$ undergo a phase transition in the holographic mutual information is a problem of finding and comparing extremal surfaces in $\mathbb{H}^d.$ This problem is nontrivial, as the differential equation satisfied by extremal surfaces in a generic coordinate system is a second-order, nonlinear ODE. We may simplify the problem considerably by choosing a system of coordinates that is well adapted to the regions we consider.

\begin{figure}[h]
\centering
%
%
%
\includegraphics{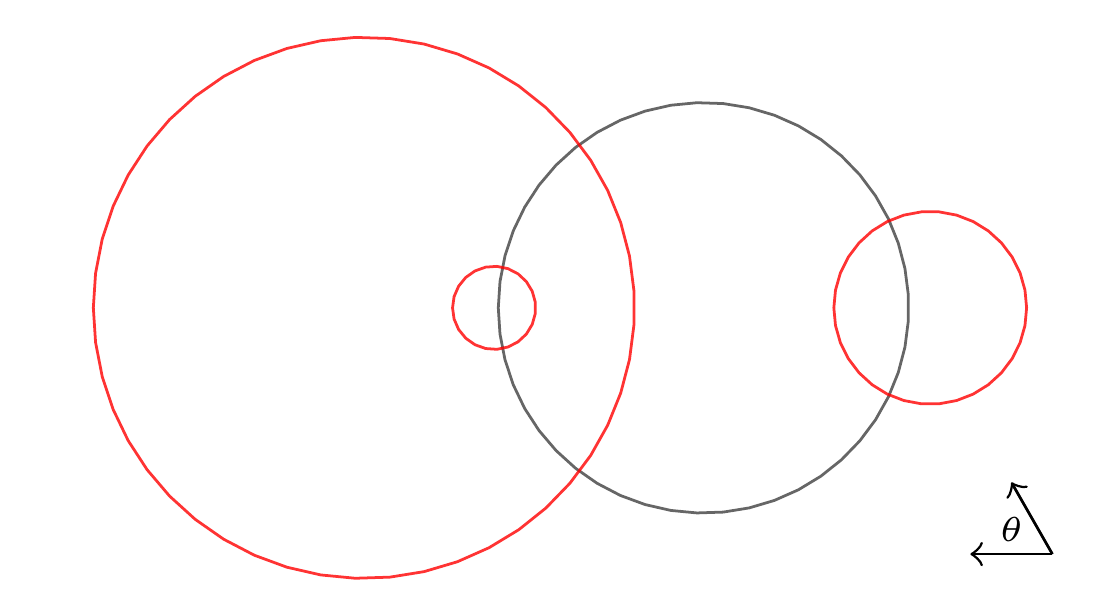}
\caption{Spheres (red) centered on the $\theta = 0$ axis that intersect the boundary of $\mathbb{H}^2$ (gray) orthogonally. The family of minimal caps obtained by restricting these spheres to the interior of $\mathbb{H}^2$ form a foliation of $\mathbb{H}^2.$ Points that are symmetric across the $\theta = \pi/2$ plane lie on different spheres with the same radius $\rho.$}
\label{fig:orthogonal-spheres}
\end{figure}

Our choice of coordinate system is motivated by the fact that one class of extremal surfaces homologous to regions of the form $A \cup B$ is known, namely the class disconnected surfaces formed by the union of two minimal caps homologous to $A$ and $B$. We will adapt our coordinates to this knowledge by choosing a coordinate that has as its level sets minimal caps with their centers at $\theta = 0$. Crucially, as we will see, the integral curves of this coordinate are isometries of hyperbolic space; this feature is what allows us to simplify our calculation. Each minimal cap is a portion of a sphere in the ambient space $0 \leq r < \infty$ with its center on the axis $\theta = 0$. Since these spheres must meet $r=1$ orthogonally, each point in $\mathbb{H}^d$ must lie on exactly one such sphere. We will naively choose one of our coordinates to be the radius of the corresponding sphere, $\rho$. Several such spheres are sketched in $d=2$ in Figure \ref{fig:orthogonal-spheres}. The condition that the sphere meets $r=1$ orthogonally constrains $\rho$ to satisfy
\begin{equation} \label{eq:rho}
1+ \rho^2 = \frac{1 + r^2}{2 r |\cos{\theta}|}
\end{equation}
As a coordinate, $\rho$ can only cover one half of $\mathbb{H}^d$ at a time, since points in $\mathbb{H}^d$ that are reflection-symmetric across the plane $\theta = \pi/2$ have the same value of $\rho$ (cf.\ Figure \ref{fig:orthogonal-spheres}).

From equation \eqref{eq:rho}, it is straightforward to check that surfaces of constant $\rho$ are extremal surfaces, as previously claimed. One simply defines a normal vector to any such surface as $n^a = \Del^a \rho$, normalizes it to $\hat{n}^a = n^a / \sqrt{n^b n_b}$, then checks that the unit normal vector satisfies
\begin{equation}
	\Del_a \hat{n}^a = 0
\end{equation}
on the surface, which is a necessary and sufficient condition for a surface to be locally extremal \cite{cutoffs}. The fact that surfaces of constant $\rho$ are not only extremal, but form a foliation of $\mathbb{H}^d$ by extremal surfaces, suggests that there exists an isometry which maps between different minimal caps. In other words, it suggests that there exists a vector field proportional to $n^a$ that is a symmetry of the metric. By reparametrizing $\rho$ according to
\begin{equation} \label{eq:eta}
	\tanh\eta = \frac{1}{1 + \rho^2} = \frac{2 r |\cos{\theta}|}{1 + r^2},
\end{equation}
we can make this isometry explicit. It is straightforward to show that the coordinate vector field $(\del / \del \eta)^a$ is in fact a Killing vector field, i.e., it satisfies
\begin{equation}
	\Del_a (\del / \del \eta)_b + \Del_b (\del / \del \eta)_a = 0.
\end{equation}
Since $(\del / \del \eta)_a$ is a symmetry of the metric, the metric components in a system of coordinates including $\eta$ will have no $\eta$-dependence. This is the fundamental simplification that allows us to find the desired extremal surfaces numerically in any dimension and analytically in the limit of arbitrarily many dimensions. Naively, the coordinate $\eta$ ranges from zero at $\theta = \pi/2$ to $\infty$ at $\theta = 0$ --- however, since $\tanh(\eta)$ is an odd function of $\eta$ and $\cos(\theta)$ is odd under reflections about $\pi/2$, the coordinate range may be extended to $\eta \in (-\infty, \infty)$ so that $\eta$, unlike $\rho$, covers the entire geometry $\mathbb{H}^d.$

\begin{figure}[h]
\centering
\subfloat[\label{fig:constant-eta}]{
%
%
\includegraphics{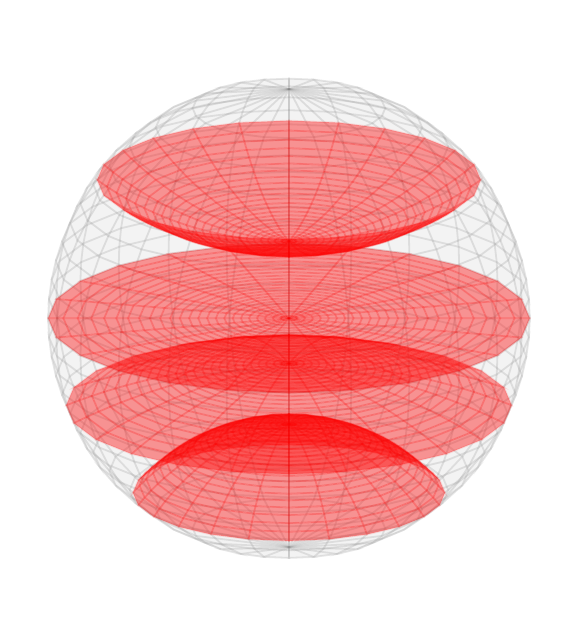}
}
\hspace{0.5cm}
\subfloat[\label{fig:constant-zeta}]{
%
%
\includegraphics{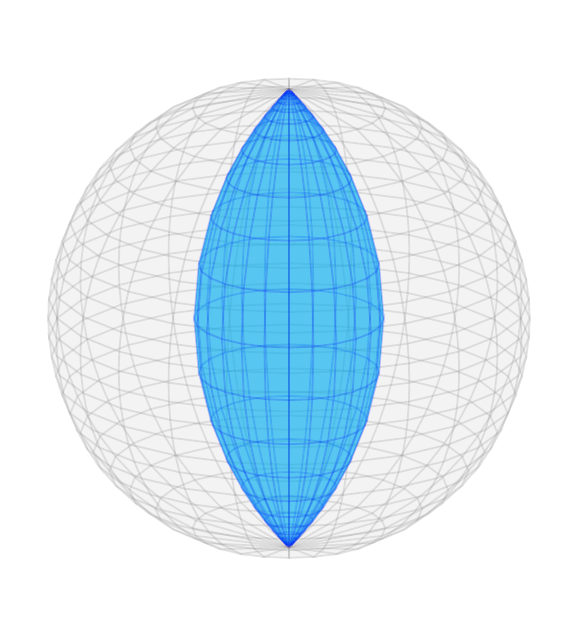}
}
	\caption{Surfaces of (a) constant $\eta$ and (b) constant $\zeta$ in the coordinate system for $\mathbb{H}^3$ given by equations \eqref{eq:r-transform} and \eqref{eq:theta-transform}. The surface plotted are (a) $\eta = 0.7, 0, -0.4, -1$ and (b) $\zeta = 0.75$.}
	\label{fig:level-surfaces}
\end{figure}

All that remains is to adapt the other coordinates on the Poincar\'{e} ball to $\eta.$ Surfaces of constant $\eta$ are symmetric under rotations along the $(d-2)$-sphere, so we need only find one other compatible coordinate, which we call $\zeta$, in order to implement the coordinate transform $(r, \theta) \mapsto (\eta, \zeta).$ For convenience, we choose $\zeta$ such that level sets of $\zeta$ are (i) orthogonal to level sets of $\eta$, and (ii) symmetric under rotations along the $(d-2)$-sphere. The level sets of $\zeta$ will then be surfaces of revolution around the $\theta=0$ axis whose cross-sections are segments of circles that intersect the boundary at $\theta=0$ and $\theta = \pi$ --- see Figure \ref{fig:constant-zeta} for a sketch. One convenient choice of coordinate satisfying these criteria is to let $\zeta$ be the angle formed between the $\theta=0$ axis and the $\zeta=\mathrm{const.}$ surface at the point on the boundary where the two meet. The function $\zeta$ then satisfies
\begin{equation} \label{eq:zeta}
	\cos{\zeta} = \frac{1-r^2}{\sqrt{(1+r^2)^2 - 4 r^2 \cos^2{\theta}}}.
\end{equation}

The combined coordinate transform $(r, \theta) \mapsto (\eta, \zeta)$ can be found by inverting equations \eqref{eq:eta} and \eqref{eq:zeta}, and is given by
\begin{eqnarray}
	r^2 
		& = & \frac{\sin^2{\zeta} + \sinh^2{\eta}}{(\cos{\zeta} + \cosh{\eta})^2}, \label{eq:r-transform} \\
	\tan{\theta}
		& = & \frac{\sin{\zeta}}{\sinh{\eta}}. \label{eq:theta-transform}
\end{eqnarray}
In these coordinates, the metric of $\mathbb{H}^d$ given in equation \eqref{eq:poincare-ball} takes the form
\begin{equation} \label{eq:bispherical-metric}
	ds^2 = \frac{1}{\cos^2{\zeta}} \left( d \eta^2 + d \zeta^2 + \sin^2{\zeta}\, d \Omega^2_{d-2} \right).
\end{equation}
The coordinate $\eta$ ranges from $+\infty$ at the north pole of the sphere, to zero at the plane $\theta = \pi / 2$, to $-\infty$ at the south pole. The coordinate $\zeta$ ranges from zero at the $\theta = 0$ axis to $\pi / 2$ at the boundary. To help with visualization, some level sets of $\eta$ and $\zeta$ are sketched in $d=3$ in Figure \ref{fig:level-surfaces}.

\subsection{Mutual Information at Large $d$}
\label{sec:large-d-MI}

When $A$ and $B$ are sufficiently large, there are two different classes of extremal surfaces homologous to $A \cup B$. One surface, $\Sigma_{\mathrm{cap}}$, is just the union of the minimal caps homologous to $A$ and $B$ individually. There may also exist connected, ``tube-like'' extremal surfaces like the one sketched in Figure \ref{fig:connected-min-surface}. In general, no such tube surface will exist when $A$ and $B$ are sufficiently small, and when $A$ and $B$ are large there exist two tube-like surfaces that extremize the area action; we show the existence of these two branches of ``tube-like'' solutions numerically in subsection \ref{sec:numerics} (cf. Figure \ref{fig:etas-plot}). \footnote{\ 
While two branches of smooth ``tube-like'' surfaces exist only in dimensions $d \geq 3$, the presence of these branches can be seen in $d=2$. There are in fact two ``connected'' minimal surfaces linking antipodal intervals in $d=2$: the usual one consisting of two smooth geodesics (cf.\ Figure \ref{f:AdS3}), and also the ``kinked surface'' consisting of piecewise-smooth geodesics that touch at the center of $\mathbb{H}^2$ (cf.\ Figure \ref{fig:both-branches}).} The mutual information phase transition occurs when $A$ and $B$ are sufficiently large that one of the tube-like surfaces becomes smaller than the ``cap'' surface in the sense that its finite area difference with the cap surface becomes negative.

Both the ``cap'' and ``tube'' surfaces are symmetric under rotations along the $(d-2)$-sphere, so both locally satisfy equations of the form
\begin{equation}
	\eta
		= \eta(\zeta).
\end{equation}
On such a surface, the induced metric $h_{ab}$ inherited from \eqref{eq:bispherical-metric} takes the form
\begin{equation}
	ds^2
		= \frac{1}{\cos^2{\zeta}} ((1 + \dot{\eta}^2) d \zeta^2 + \sin^2{\zeta} d\Omega^2_{d-2}),
\end{equation}
where the overdot denotes a total derivative with respect to $\zeta.$ The volume element of the induced metric is given by
\begin{equation} 
	\sqrt{h}
		= \sec{\zeta} \tan^{d-2}\zeta \sqrt{1 + \dot{\eta}^2} \sqrt{\omega},
\end{equation}
where $\sqrt{\omega}$ is the volume element of $d \Omega^2_{d-2}$. For notational convenience, we denote by $f(\zeta)$ the function
\begin{equation} \label{eq:f}
	f(\zeta) \equiv \sec{\zeta} \tan^{d-2}\zeta,
\end{equation}
so the volume element takes the form
\begin{equation} \label{eq:volume-element}
	\sqrt{h}
		= f(\zeta) \sqrt{1 + \dot{\eta}^2} \sqrt{\omega}.
\end{equation}

\begin{figure}
\centering
%
%
%
\includegraphics{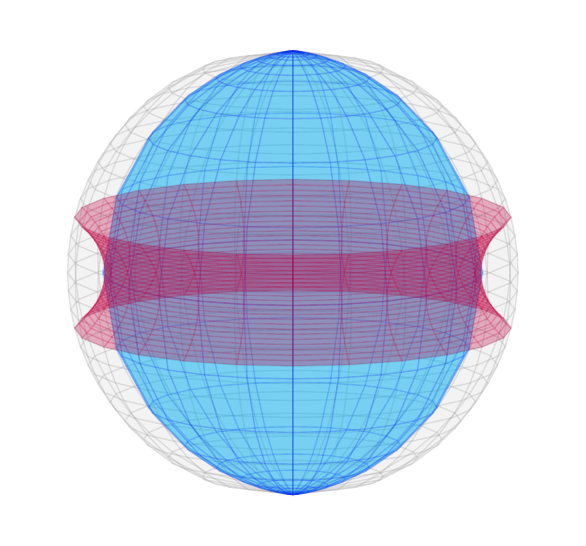}
\caption{The tube-like surface (purple) from Figure \ref{fig:connected-min-surface} together with the surface of constant $\zeta$ (blue) that marks its ``turning point'', $\zeta_*$. This particular surface has turning point $\zeta_* \sim 0.93$.}
\label{fig:turning-point}
\end{figure}

The Euler-Lagrange equations of this volume element are given by
\begin{equation} \label{eq:euler-lagrange}
	\dot{\eta}^2 = \frac{C^2}{f(\zeta)^2 - C^2},
\end{equation}
where $C$ is a constant. One possible solution to this equation is given by $C = 0$, so that $\dot{\eta} = 0.$ These are the minimal caps that are symmetric about the $\zeta = 0$ axis, which we showed in the previous subsection are surfaces of constant $\eta.$ The (infinite) area of any such surface $\Sigma_{\mathrm{cap}}$ is given formally according to equation \eqref{eq:volume-element} by
\begin{equation}
	\mathcal{A}_{\mathrm{cap}}
		= \int_{\Sigma_{\mathrm{cap}}} \sqrt{h} = S_{d-2} \int_0^{\pi/2} d\zeta\, f(\zeta),
\end{equation}
where $S_{d-2}$ is the surface area of the $(d-2)$-sphere. The other class of solutions to \eqref{eq:euler-lagrange}, with $C \neq 0$, are the ``tube-like'' solutions. These solutions have a special point, which we call $\zeta_*$, where $C^2 = f(\zeta_*)^2$ is satisfied and the derivative $\dot{\eta}$ blows up. This point corresponds to the ``turning point'' of the tube in the bulk --- the minimal value of $\zeta$ reached by the tube as it extends from $(A \cup B)^c$ into the bulk. A tube-like surface with its turning point identified is sketched in Figure \ref{fig:turning-point}. The (infinite) area of any such surface $\Sigma_{\mathrm{tube}}$ is given formally by
\begin{equation}
	\mathcal{A}_{\mathrm{tube}}
		= \int_{\Sigma_{\mathrm{tube}}} \sqrt{h} = 2 S_{d-2} \int_{\zeta_{*}}^{\pi/2} d\zeta\, \frac{f(\zeta)^2}{\sqrt{f(\zeta)^2 - f(\zeta_*)^2}},
\end{equation}
where we have used $C = f(\zeta_*)$ and the factor of $2$ comes from the fact that the tube is symmetric under the reflection $\eta \mapsto -\eta$.

A choice of turning point $\zeta_*$ uniquely specifies an extremal, tube-like surface homologous to $A \cup B$. At the phase transition, there exists some $\zeta_*$ whose corresponding tube-like surface has area equal to the area of two minimal caps. At the phase transition, therefore, $\zeta_*$ satisfies
\begin{equation} \label{eq:MI-sum}
	0
		= 2 \mathcal{A}_{\mathrm{cap}} - \mathcal{A}_{\mathrm{tube}} 
		= 2 S_{d-2} \left[ \int_0^{\pi/2} d\zeta\, f(\zeta) - \int_{\zeta_*}^{\pi/2} d\zeta\, \frac{f(\zeta)^2}{\sqrt{f(\zeta)^2 - f(\zeta_*)^2}} \right].
\end{equation}
Note that since the area $S_{d-2}$ vanishes in the limit $d \rightarrow \infty$, equation \eqref{eq:MI-sum} always vanishes in the large-$d$ limit for \emph{any} configuration of boundary regions. However, for any \emph{fixed}, large $d$, there will still be a phase transition at which equation \eqref{eq:MI-sum} is satisfied. The parameters of this phase transition can be derived by dividing the above expression by $2 S_{d-2}$ and finding the value of $\zeta_*$ that solves the equation
\begin{equation} \label{eq:MI-integrals}
	0
		= \int_0^{\pi/2} d\zeta\, f(\zeta) - \int_{\zeta_*}^{\pi/2} d\zeta\, \frac{f(\zeta)^2}{\sqrt{f(\zeta)^2 - f(\zeta_*)^2}},
\end{equation}
should such a solution exist. Equivalently, one could define a \emph{normalized} mutual information $\hat{I}(A:B) = I(A:B) / S_{d-2}$, and look for the phase transition in this manifestly finite quantity. 

Both integrals in equation \eqref{eq:MI-integrals} are divergent at $\zeta = \pi/2.$ This is a manifestation of the fact that the areas of boundary-anchored extremal surfaces diverge near the boundary of AdS. However, the area \emph{difference} that appears in equation \eqref{eq:MI-integrals} is finite. It may be computed by introducing a cutoff at $\zeta_c = \pi/2 - \epsilon$ and taking the limit $\epsilon \rightarrow 0$; the answer obtained is independent of the cutoff procedure \cite{cutoffs}. Using such a cutoff procedure, we are free to split the infinite integral $\int_0^{\pi/2} d\zeta\, f(\zeta)$ into a sum of integrals over the ranges $(0, \zeta_*)$ and $(\zeta_*, \pi/2)$ and combine divergent terms to rewrite equation \eqref{eq:MI-integrals} as
\begin{equation} \label{eq:MI-integrals-2}
	0
		= \int_0^{\zeta_*} d\zeta\, f(\zeta) + \int_{\zeta_*}^{\pi/2} d\zeta\, f(\zeta) \left( 1 - \frac{f(\zeta)}{\sqrt{f(\zeta)^2 - f(\zeta_*)^2}} \right).
\end{equation}
Both integrals in equation \eqref{eq:MI-integrals-2} are now manifestly convergent.

Recall that $f(\zeta) = \sec{\zeta} \tan^{d-2}{\zeta}$ is dimension-dependent. For the special case $d=2$, the second integral in \eqref{eq:MI-integrals-2} is elementary, and equation \eqref{eq:MI-integrals-2} is solved by the critical value $\zeta_*^c = \pi/4$; this is why the mutual information phase transition in $d=2$ is straightforward to compute, as explained in the introduction to this paper. For $d=3$, the integral is elliptic; it has a closed form in terms of functions with known series expansions, which can then be exploited to find the parameters of the mutual information phase transition numerically with high accuracy.\footnote{\ 
A related elliptic integral was computed in \cite{krtous} to obtain analytic expressions for the tube-like extremal surfaces in $d=3$.} For $d > 3$, the second integral in equation \eqref{eq:MI-integrals-2} is \emph{hyperelliptic}; it is not analytically tractable, and cannot be exploited to find the value of $\zeta_*$ at which the phase transition occurs. However, both integrals in equation \eqref{eq:MI-integrals-2} are tractable in the limit $d \rightarrow \infty$ as power series in $1/(d-2)$. We will now exploit this simplification to find the limiting behavior of the phase transition as $d \rightarrow \infty$, and use this result to glean information about the dimension-dependence of holographic correlations.

We will evaluate the two convergent integrals in \eqref{eq:MI-integrals-2} separately, referring to them as $I_1$ and $I_2$, respectively. It is first convenient to remove the $\zeta_*$ dependence in the range of integration by making the coordinate substitution
\begin{equation}
	u = \frac{\tan{\zeta}}{\tan{\zeta_*}}.
\end{equation}
Under this substitution, the integrals take the form
\begin{align}
	I_1
		& = \tau^{d-1} \int_0^{1} du\, \frac{u^{d-2}}{\sqrt{1+\tau^2 u^2}}, \\
	I_2
		& = \tau^{d-1} \int_1^{\infty} du\, \frac{u^{d-2}}{\sqrt{1+\tau^2 u^2}}
				\left( 1 - \frac{\sqrt{1 + \tau^2 u^2} u^{d-2}}{\sqrt{(1 + \tau^2 u^2) u^{2(d-2)} - (1 + \tau^2)}} \right),
\end{align}
where we have written $\tau = \tan{\zeta_*}$ and used the form of $f(\zeta)$ given in equation \eqref{eq:f}. The term $u^{d-2}$ which appears in both integrands has an essential singularity as $d\rightarrow \infty.$ To facilitate a large-$d$ expansion, it is therefore convenient to make the substitution $w = u^{d-2}$, after which the integrals take the form
\begin{align}
	I_1
		& = \frac{\tau^{d-1}}{d-2} \int_0^{1} dw\, \frac{w^{1/(d-2)}}{\sqrt{1+\tau^2 w^{2/(d-2)}}}, \\
	I_2
		& = \frac{\tau^{d-1}}{d-2} \int_1^{\infty} dw\, \frac{w^{1/(d-2)}}{\sqrt{1+\tau^2 w^{2/(d-2)}}}
				\left( 1 - \frac{w \sqrt{1 + \tau^2 w^{2/(d-2)}}}{\sqrt{(1 + \tau^2 w^{2/(d-2)}) w^2 - (1 + \tau^2)}} \right).
\end{align}
The integrands can be expanded in powers of $1/(d-2)$ and evaluated analytically order-by-order. Up to the addition terms of order $O(1/(d-2)^3),$ the integrals are given by
\begin{align}
	I_1
		& = \frac{\tau^{d-1}}{d-2} \left[ \frac{1}{\sqrt{1 + \tau^2}} - \frac{1}{d-2} \frac{1}{(1+\tau^2)^{3/2}} \right], \\
	I_2
		& = \frac{\tau^{d-1}}{d-2}  \left[ - \frac{1}{\sqrt{1 + \tau^2}}
			+ \frac{1}{d-2} \frac{1}{(1 + \tau^2)^{3/2}} \left( 1 - \frac{\pi}{2}(1 - \tau^2) \right) \right].
\end{align}
Their sum satisfies
\begin{equation} \label{eq:integral-sum}
	I_1 + I_2
		= \frac{\pi}{2} \frac{\tau^{d-1}}{(d-2)^2} \frac{\tau^2 - 1}{(1 + \tau^2)^{3/2}} + O\left( \frac{1}{(d-2)^3} \right).
\end{equation}

From equation \eqref{eq:integral-sum}, we see that the area difference between the cap and tube surfaces vanishes at leading order in the large-$d$ expansion exactly when $\tau^2 = 1$. Since $\tau$ is given by $\tau = \tan{\zeta_*}$, it follows that the tube-like extremal surface that initiates the phase transition tends to
\begin{equation} \label{eq:large-d-critical-zeta}
	\lim_{d\rightarrow\infty} \zeta_*^c = \frac{\pi}{4}.
\end{equation}
The value of $\eta$ at which this surface meets the boundary, $\eta_\infty$, is computed by the integral
\begin{equation}
	\eta_{\infty}
		= \int_{\zeta_*}^{\pi/2} d\zeta\, \dot{\eta}
		= \int_{\zeta_*}^{\pi/2} d\zeta\, \frac{f(\zeta_*)}{\sqrt{f(\zeta)^2 - f(\zeta_*)^2}},
\end{equation}
where we have used the form of $\dot{\eta}$ given in equation \eqref{eq:euler-lagrange}, and used the tube-like surface's $\eta \mapsto -\eta$ symmetry to guarantee that $\eta(\zeta_*)$ vanishes. Using the same substitutions we used previously, $u = \tan\zeta / \tan{\zeta_*}$ and $w = u^{d-2}$, this integral takes the form
\begin{equation} \label{eq:eta-infty-integral}
	\eta_{\infty}
		= \frac{\tau \sqrt{1+\tau^2}}{d-2} \int_{1}^{\infty} dw\,  \frac{w^{1/(d-2)}}{w(1+w^{2/(d-2)} \tau^2)} \frac{1}{\sqrt{(1+\tau^2 w^{2/(d-2)}) w^2 -  (1+\tau^2)}}.
\end{equation}
Since the leading term in the integrand is $O(1)$ in $1/(d-2)$, we see immediately that $\eta_\infty$ falls off as $1/(d-2)$ in the limit of large $d$. In other words, \emph{the boundary separation at which the phase transition occurs approaches zero as $1/d$}. The ball-shaped boundary regions $A$ and $B$ must have entangling surfaces within a $(1/d)$-size neighborhood of the equator in order to develop nonzero correlations at order $G_N^{-1}.$ We can compute the $d$-dependence of $\eta$ explicitly by performing the integral in equation \eqref{eq:eta-infty-integral} at leading order in $1/(d-2)$, yielding the expression
\begin{align}
	\eta_{\infty}
		& = \frac{1}{d-2} \frac{\tau}{1+\tau^2} \int_{1}^{\infty} dw\,  \frac{1}{w \sqrt{w^2-1}} + O\left(\frac{1}{(d-2)^2}\right) \nonumber \\
		& = \frac{1}{d-2} \frac{\pi}{2} \frac{\tau}{1+\tau^2} + O\left(\frac{1}{(d-2)^2}\right).
\end{align}
Substituting in $\tau = \tan\zeta_*$ gives $\eta_\infty$ in a simple closed form:
\begin{equation} \label{eq:eta-infty}
	\eta_\infty
		= \frac{1}{d-2} \frac{\pi}{2} \sin{\zeta_*} \cos{\zeta_*} + O\left(\frac{1}{(d-2)^2}\right).
\end{equation}
We argued above that the position of the phase transition is given to leading order in $1/(d-2)$ by $\zeta_*^c = \pi/4$, yielding a critical value of $\eta_\infty$ given by
\begin{equation} \label{eq:eta-infty-star}
	\eta_\infty^c
		= \frac{1}{d-2} \frac{\pi}{4} + O\left(\frac{1}{(d-2)^2}\right).
\end{equation}
By transforming back to the usual spherical coordinates using equation \eqref{eq:theta-transform}, we see that the critical angle at which antipodal caps develop correlations at order $G_N^{-1}$ limits to
\begin{equation}
	\theta_\infty^c = \frac{\pi}{2} - \frac{1}{d-2} \frac{\pi}{4} + O\left(\frac{1}{(d-2)^2}\right).
\end{equation}
Notably, this implies that boundary caps must occupy $100\%$ of the area of the boundary with respect to the spherically symmetric metric to develop large correlations in the limit $d \rightarrow \infty$; in order for the strip between the two critical caps to occupy a nonvanishing fraction of the boundary area, the critical angle would have to differ from $\pi/2$ at order $1/\sqrt{d-2}.$\footnote{\
\label{fn:sqrt-d-scaling}A $\Delta \theta$-width strip around the equator of a $(d-1)$-sphere occupies fractional area \begin{equation}\frac{1}{\sqrt{\pi}} \frac{\Gamma(d/2)}{\Gamma((d-1)/2)} \int_{-\Delta \theta/2}^{\Delta \theta/2} d\theta\, \cos^{d-2}(\theta).\end{equation} In the large-$d$ limit, this expression approaches
\begin{equation} \sqrt{\frac{d}{2 \pi}} \int_{-\Delta \theta/2}^{\Delta \theta/2} d\theta\, e^{- (d-2) \theta^2 /2}.\end{equation} Standard estimates on the error of Gaussian integrals show that this expression limits to zero when $\theta$ vanishes more quickly than $1/\sqrt{d-2}$ and to one when $\theta$ vanishes more slowly than $1/\sqrt{d-2}.$} While these expressions for $\eta_\infty^c$ and $\theta_\infty^c$ are coordinate dependent, we will re-express the boundary separation at which the phase transition occurs in terms of a coordinate-independent and conformally invariant quantity in subsection \ref{sec:conformal}.

One interesting feature of this calculation is that the expression for $\eta_\infty$ in terms of $\zeta_*$ given by equation \eqref{eq:eta-infty} is general --- it holds for any tube-like surface, not just the one that initiates the mutual information phase transition. We see by inspection of equation \eqref{eq:eta-infty} that $\eta_{\infty}$ is \emph{bounded} at leading order as a function of $\zeta_*$. In other words, $\eta_{\infty}$ cannot be arbitrarily large; in fact, it cannot be larger than order $1/(d-2).$ The fact that $\eta_\infty$ has a maximal value suggests that tube-like extremal surfaces do not exist for arbitrary values of $\eta_\infty$ in large dimensions --- there is a critical value of $\eta_\infty$ near the equator where tube-like extremal surfaces begin to exist. We also see that once such solutions exist, there generically exist two values of $\zeta_*$ satisfying equation \eqref{eq:eta-infty} for a single value of $\eta_\infty$; i.e., there are two tube-like extremal surfaces corresponding to any permitted boundary condition $\eta_\infty$ (except for the maximal allowed value of $\eta_\infty$, at which only one such surface exists). In the following subsection, we show these properties explicitly for dimensions $d=3$ through $d=20$ using numerical techniques. By inspection of equation \eqref{eq:eta-infty}, we see that the maximal value of $\eta_\infty$ is attained in the large-$d$ limit when $\zeta_* = \pi/4$ --- in other words, the critical value of $\eta_\infty$ at which tube-like extremal surfaces begin to exist and the critical value of $\eta_\infty$ at which the mutual information phase transition occurs converge at leading order in the limit $d \rightarrow \infty.$

Finally, we note that equation \eqref{eq:integral-sum} can be expanded analytically beyond leading order in $1/(d-2)$ to obtain approximations for the critical parameter $\zeta_*^c$ in finite dimensions. Expanding to one further order in $1/(d-2)$, the value of $\zeta_*^c$ predicted by equation \eqref{eq:integral-sum} for $d \geq 5$ is within $4\%$ percentage error of the numerical values we compute in the following subsection. This suggests that the $1/(d-2)$ expansion is genuinely computationally useful for studying holographic entanglement entropy in physically interesting dimensions, just as it was found to be computationally useful for studying black holes in general relativity in \cite{largeD}. We comment further on this calculation in the discussion in Section \ref{sec:discussion}.

\subsection{Numerical Results in Intermediate Dimensions}
\label{sec:numerics}

With the parameters of the mutual information phase transition computed analytically at large $d$, we now turn to computing the parameters numerically for intermediate dimensions $d=3$ through $d=20$ where analytic calculations are intractable. Recall from equation \eqref{eq:MI-integrals-2} that the mutual information phase transition is characterized by the turning point $\zeta_*$ that satisfies equation
\begin{equation} \label{eq:MI-integrals-revisited}
	0
		= \int_0^{\zeta_*} d\zeta\, f(\zeta) + \int_{\zeta_*}^{\pi/2} d\zeta\, f(\zeta) \left( 1 - \frac{f(\zeta)}{\sqrt{f(\zeta)^2 - f(\zeta_*)^2}} \right),
\end{equation}
with $f(\zeta)$ given by
\begin{equation}
	f(\zeta) \equiv \sec{\zeta} \tan^{d-2}\zeta.
\end{equation}
In principle, one can simply compute these two integrals numerically for various values of $\zeta_*$ and search for values of $\zeta_*$ for which equation \eqref{eq:MI-integrals-revisited} nearly vanishes. In practice, most numerical algorithms will fail to converge for sufficiently large $d$, since evaluating the integrand of the second integral requires taking the ratios and differences of rather large numbers, which can result in loss of machine precision. To avoid this complication, it is useful to make the substitution
\begin{equation}
	h
		= \sqrt{\frac{f(\zeta) - f(\zeta_*)}{f(\zeta) + f(\zeta_*)}},
\end{equation}
in which case equation \eqref{eq:MI-integrals-revisited} may be expressed as
\begin{equation}
	0
		= \int_0^{\zeta_*} d\zeta\, f(\zeta) - \int_{\zeta_*}^{\pi/2} d\zeta\, f(\zeta_*) \frac{(1+h^2)(1-h)}{2 h (1+h)}.
\end{equation}
It is also useful to explicitly factor out the term $f(\zeta_*)$, since this quantity has a pole near $\zeta_* = \pi/2.$ The final expression for the integral is
\begin{equation} \label{eq:MI-numerics}
	0
		= f(\zeta_*) \left[ \int_0^{\zeta_*} d\zeta\, \frac{f(\zeta)}{f(\zeta_*)} - \int_{\zeta_*}^{\pi/2} d\zeta\, \frac{(1+h^2)(1-h)}{2 h (1+h)} \right].
\end{equation}

\begin{figure}[h]
	\centering
	\subfloat[\label{fig:MI-plot}]{
	\makebox[\textwidth][c]{\includegraphics[scale=0.285]{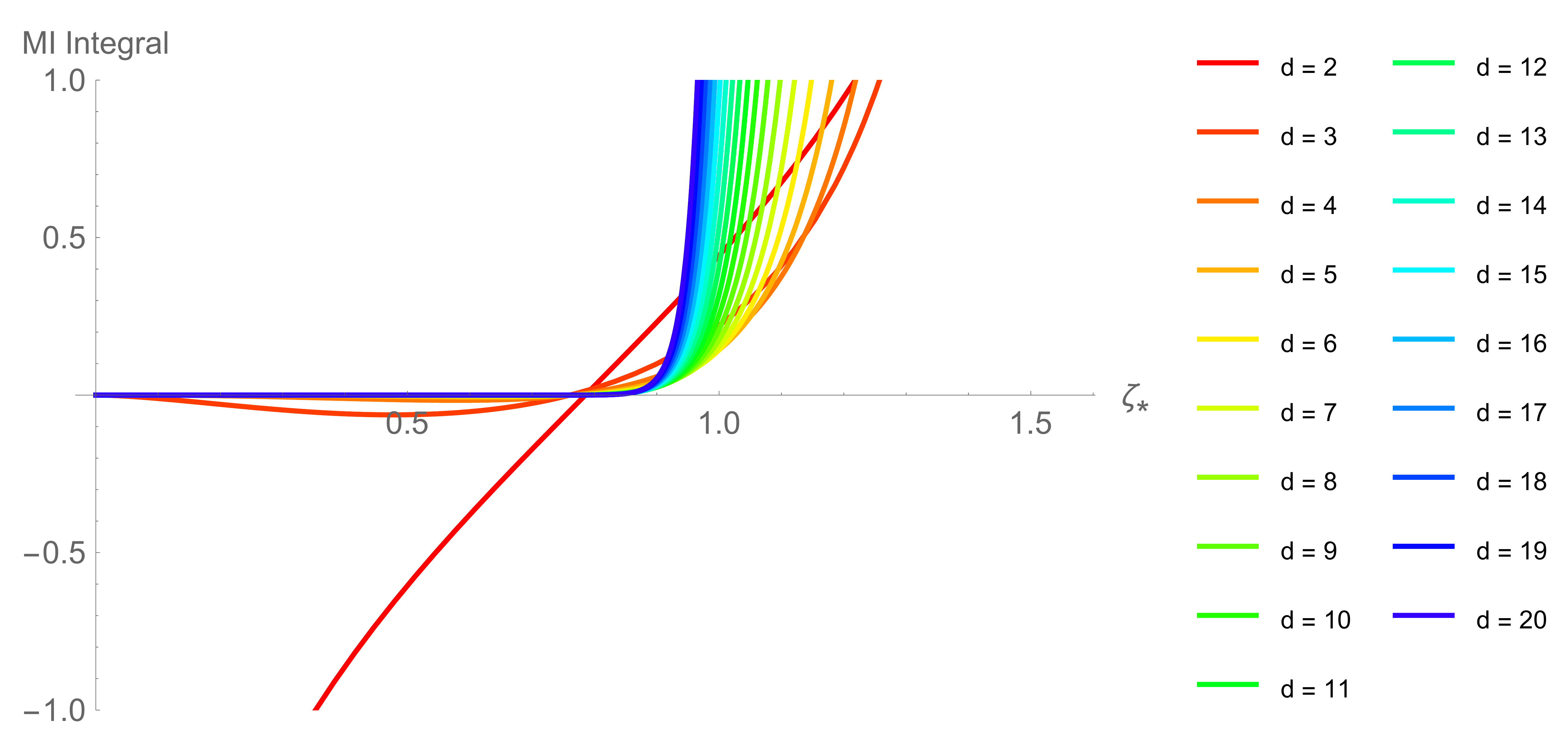}}
	}
	\hspace{0.5cm}
	\subfloat[\label{fig:MI-plot-zoom}]{
	\makebox[\textwidth][c]{\includegraphics[scale=.285]{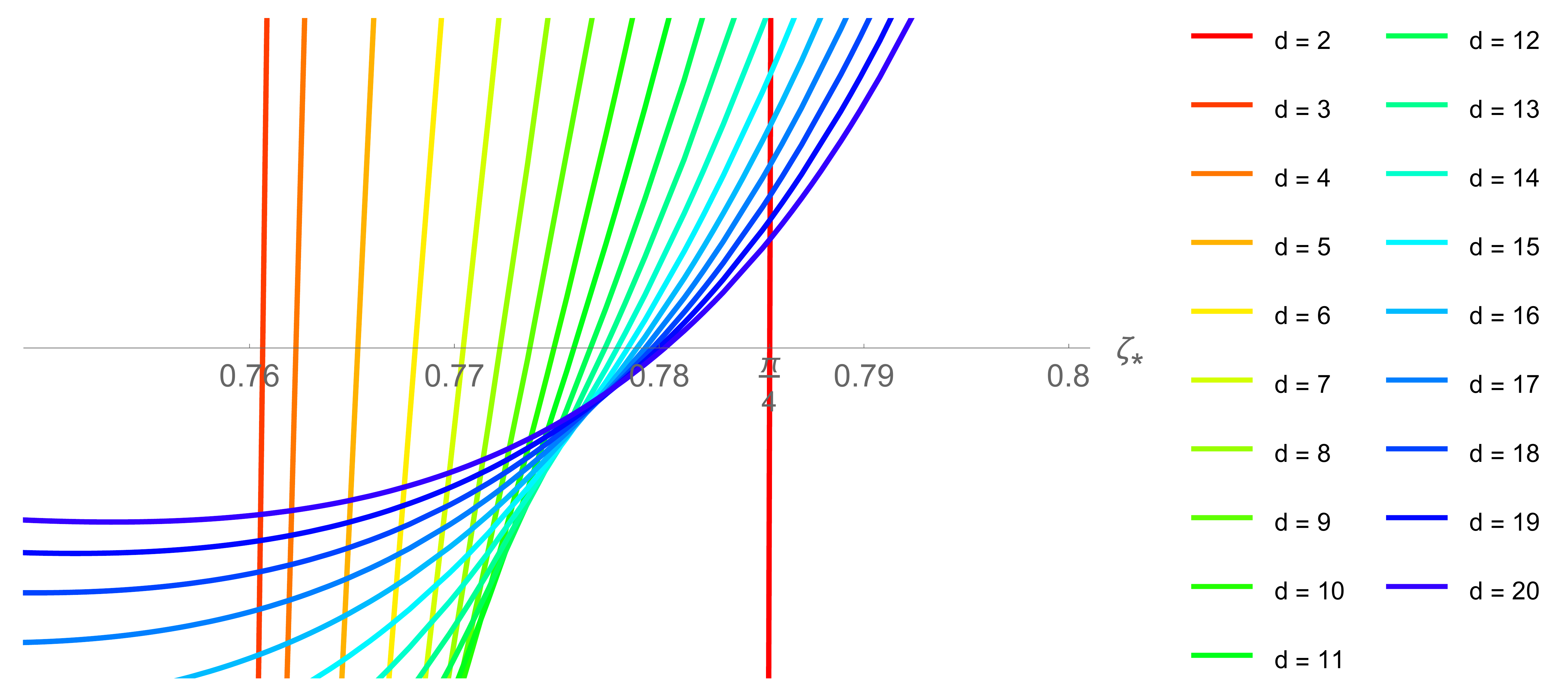}}
	}
	\caption{(a) Plots of the right-hand side of equation \eqref{eq:MI-numerics} for all values of $d$ between $2$ and $20.$ (b) The same plot,
	zoomed in near $\zeta_* = \pi/4.$ We see that the critical value of $\zeta_*$ is exactly $\pi/4$ at $d=2$, jumps down to ${\sim}0.76$ at $d=3$, then increases monotonically toward $\pi/4$ as $d \rightarrow \infty.$}
	\label{fig:MI-plots}
\end{figure}

In Figure \ref{fig:MI-plot}, we show plots of the right-hand side of equation \eqref{eq:MI-numerics} for several values of $d$ between $d=2$ and $d=20$ for the full range of possible turning points $\zeta_*=0$ to $\zeta_* = \pi/2$. In Figure \ref{fig:MI-plot-zoom}, we show the same plots zoomed in near $\zeta_* = \pi/4$; by inspection of this plot, one can observe the qualitative behavior of the root of equation \eqref{eq:MI-numerics} for dimensions $d=2$ to $d=20.$ In dimension $d=2$, equation \eqref{eq:MI-numerics} is satisfied exactly at $\zeta_* = \pi/4$, as can be computed analytically by evaluating the elementary integrals in equation \eqref{eq:MI-numerics} for $d=2$. At $d=3,$ the phase transition parameter $\zeta_*^c$ jumps down to a value near ${\sim}0.76$, then slowly climbs back up towards $\pi/4$ as $d$ is further increased. We observe, in this limiting behavior, a numerical manifestation of the explicit large-$d$ limit obtained in equation \eqref{eq:large-d-critical-zeta}.

\begin{figure}[h]
	\centering
	\makebox[\textwidth][c]{\includegraphics[scale=0.285]{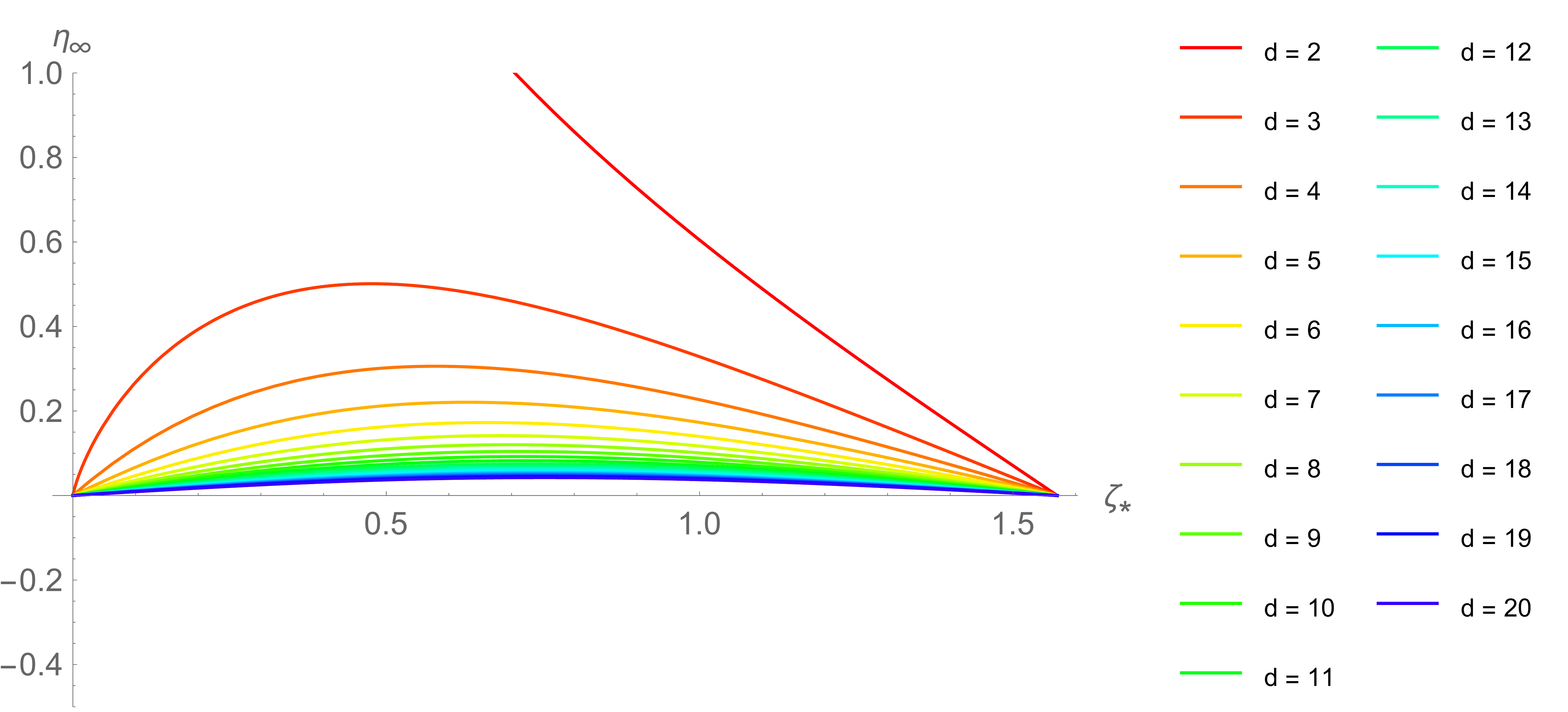}}
	\caption{The boundary condition $\eta_\infty$ as a function of the turning point $\zeta_*$ for extremal tube-like surfaces in dimensions $d=2$ through $d=20.$ While there is no maximal value of $\eta_\infty$ for $d=2$, there exists a maximal $\eta_\infty$ in $d>2$ --- above this value of $\eta_\infty$, extremal tube-like surfaces do not exist.}
	\label{fig:etas-plot}
\end{figure}

The boundary value $\eta_{\infty}$ corresponding to a particular value of the turning point $\zeta_*$ can be computed numerically by evaluating the integral
\begin{equation} \label{eq:eta-infty-numerics}
	\eta_{\infty}
		= \int_{\zeta_*}^{\pi/2} d\zeta\, \frac{f(\zeta_*)}{\sqrt{f(\zeta)^2 - f(\zeta_*)^2}}.
\end{equation}
This integral is straightforward to compute numerically. In Figure \ref{fig:etas-plot}, we show plots of $\eta_\infty$ versus $\zeta_*$ for the same values of dimension $d$ that were sketched in Figure \ref{fig:MI-plots}. We see that while there is no upper bound on $\eta_\infty$ in $d=2$, $\eta_\infty$ has a maximum value in any dimension $d > 2.$ Furthermore, the function $\eta_\infty(\zeta_*)$ is not invertible --- in $d>2$, there generically exist two tube-like surfaces corresponding to a single value of $\eta_\infty$, each with different turning points $\zeta_*.$ By comparing with the plot of the normalized mutual information function shown in Figure \ref{fig:MI-plot}, we see that the value of $\zeta_*$ for which the mutual information becomes positive is always strictly greater than the value of $\zeta_*$ corresponding to the maximal $\eta_\infty$; it is therefore always the outer branch of extremal tube-like surfaces that initiates the phase transition in the mutual information.

\begin{figure}[h]
	\centering
	\makebox[\textwidth][c]{
	\subfloat[\label{fig:stable-branch}]{
	\includegraphics[scale=0.18]{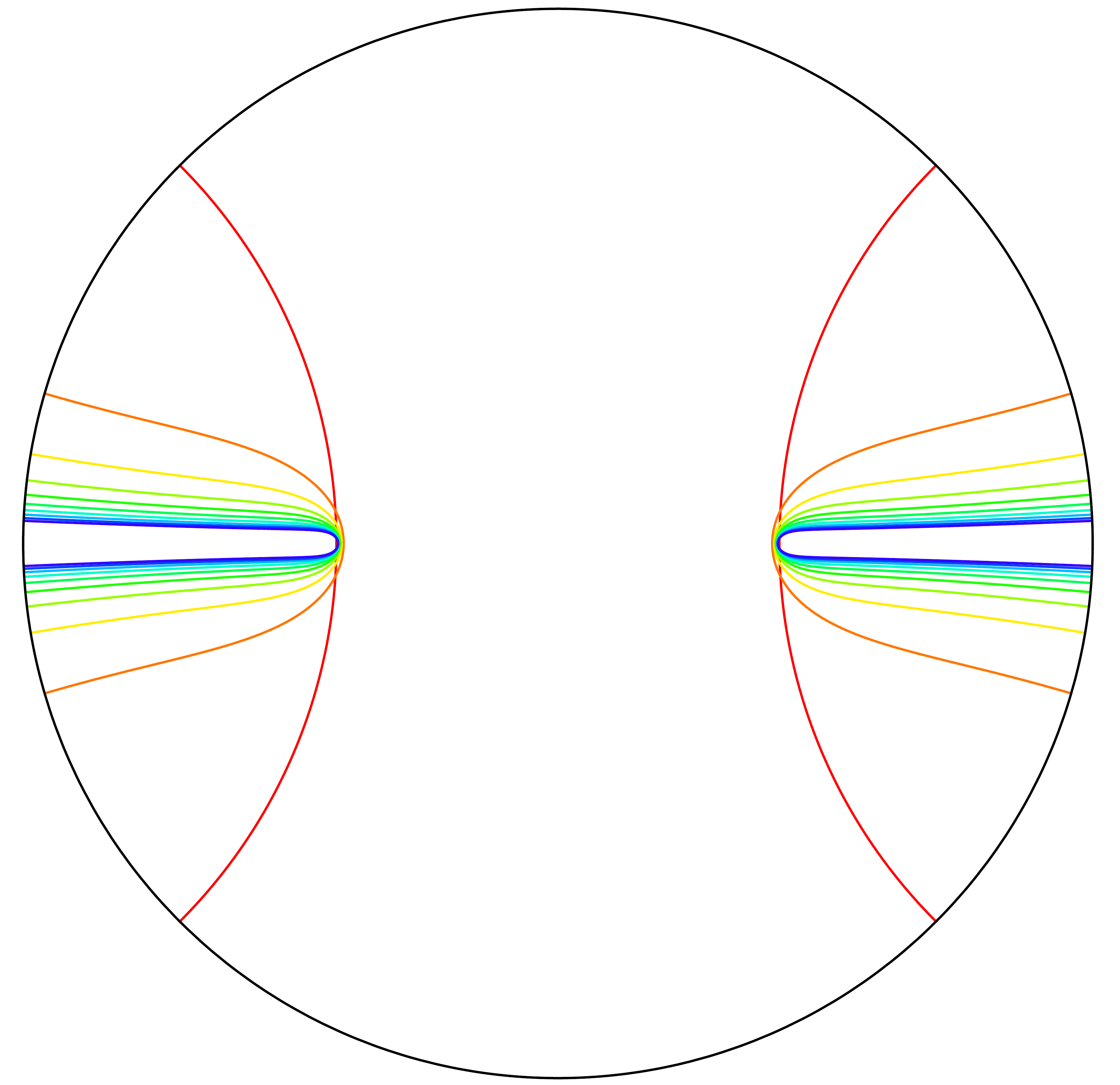}
	}
	\subfloat{
	\includegraphics[scale=0.2]{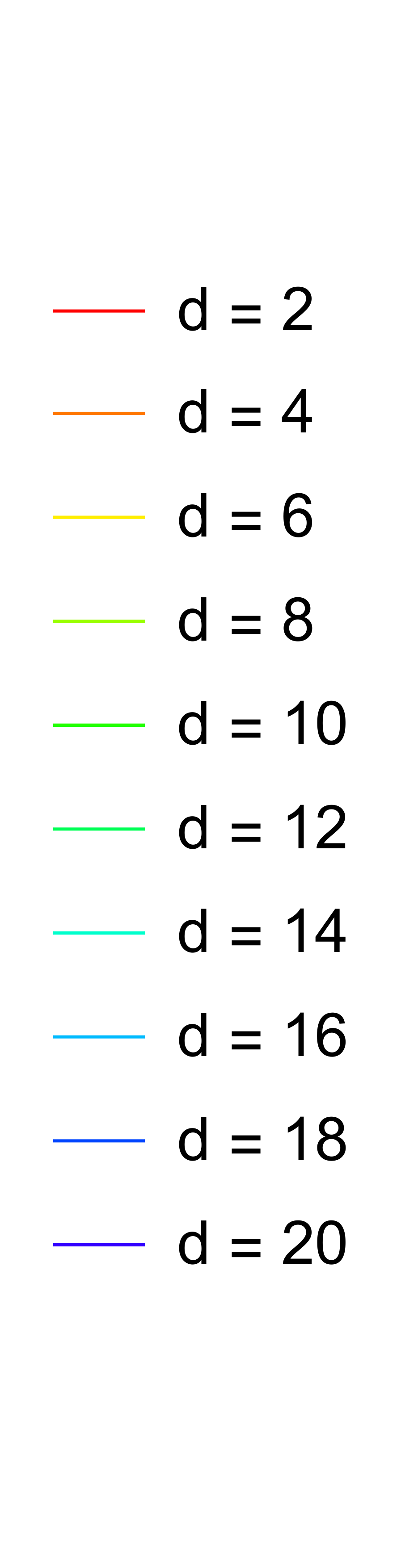}
	}
	\setcounter{subfigure}{1}
	\subfloat[\label{fig:both-branches}]{
	\includegraphics[scale=0.18]{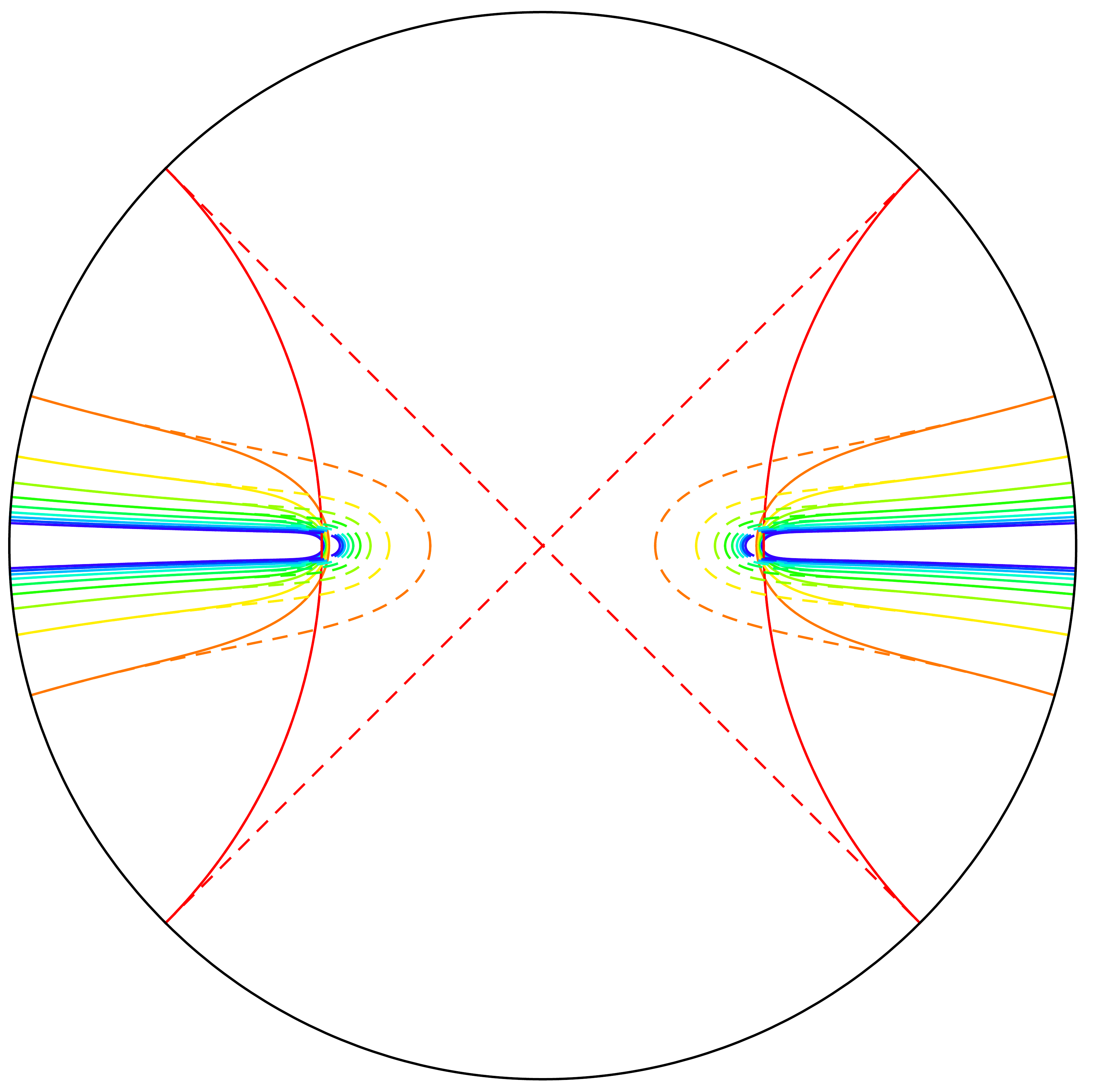}
	}
}
	\caption{(a) The tube-like minimal surfaces at the mutual information phase transition, projected down onto a single hyperbolic slice, for even dimensions in the range $d=2$ through $d=20.$ (b) The same surfaces together with their ``inner branches'' --- the other set of tube-like surfaces with the same boundary conditions --- plotted with dashed lines.}
	\label{fig:surface-branches}
\end{figure}

To aid with visualization, we have plotted in Figure \ref{fig:stable-branch} each of the ``tube-like'' surfaces that initiates the phase transition for even dimensions in the range $d=2$ through $d=20$. This is accomplished by evaluating \eqref{eq:MI-numerics} numerically to find the critical turning point $\zeta_*^c$ as a function of $d$, and then integrating
\begin{equation}
	\eta(\zeta)
		= \int_{\zeta_*^c}^{\zeta} d\tilde{\zeta}\, \frac{f(\zeta_*^c)}{\sqrt{f(\tilde{\zeta})^2 - f(\zeta_*^c)^2}}
\end{equation}
to obtain an equation for the corresponding surface. In Figure \ref{fig:stable-branch}, the surfaces are projected down onto a single slice $\mathbb{H}^2$ slice of the $d$-dimensional hyperbolic geometry so that they may be plotted together. Figure \ref{fig:both-branches} shows all of these surfaces together with their corresponding \emph{inner branches}---the other set of tube-like surfaces that have the same boundary condition $\eta_\infty$ as those that initiate the phase transition.

In future work, it may be useful to have explicit numerical values for the parameters of the phase transition in low dimensions. For this reason, table \ref{tab:numerics} contains the critical values $\zeta_*^c$ and $\eta_\infty^c = \eta_\infty(\zeta_*^c)$ that initiate the mutual information phase transition for several dimensions between $d=3$ and $d=20,$ expressed to five digits of precision. It also includes the corresponding boundary values of $\theta$ in usual spherical coordinates, computed by transforming $\eta \rightarrow \theta$ using equation \eqref{eq:theta-transform}, as well as the critical values of the conformal invariant $\chi$ we introduce in the following subsection in equation \eqref{eq:chi}.

\begin{table}
\centering
\begin{tabular}{c | c | c | c | c}
	$d$ 	& $\zeta_*^c$ 			& $\eta_\infty^c = \eta_\infty(\zeta_*^c)$  		& $\theta_{\infty}^c$ 	& $\chi^c$\\
	\hline
	$2$ 	& $\pi/4$ ($0.78540$)		& $\mathrm{arccosh}\sqrt{2}$ ($0.88137$)  	& $\pi/4$ ($0.78540$)  	& $4$\\
	$3$ 	& $0.76115$ 			& $0.43845$ 		& $1.1458$ 		& $0.81951$ \\
	$4$ 	& $0.76227$ 			& $0.28817$ 		& $1.2865$ 		& $0.34147$ \\
	$5$ 	& $0.76539$ 			& $0.21334$ 		& $1.3591$ 		& $0.18484$ \\
	$7$ 	& $0.77044$ 			& $0.13954$ 		& $1.4317$ 		& $0.07839$ \\
	$9$ 	& $0.77369$ 			& $0.10336$ 		& $1.4676$ 		& $0.04289$ \\
	$11$ 	& $0.77585$ 			& $0.08199$ 		& $1.4889$ 		& $0.02695$ \\
	$14$ 	& $0.77796$			& $0.06253$ 		& $1.5083$ 		& $0.01566$ \\
	$17$ 	& $0.77932$ 			& $0.05051$ 		& $1.5203$ 		& $0.01021$ \\
	$20$ 	& $0.78027$ 			& $0.04236$ 		& $1.5285$ 		& $0.00718$ \\
	$\infty$ & $\pi/4$ ($0.78540$)	 	& $\pi/4(d-2)$ ($0$) 		& $\pi/2 - \pi/4(d-2)$ ($0$) 		& $\pi^2/4 (d-2)^2$ ($0$)
\end{tabular}
\caption{A table containing the numerical values of the critical parameters $\zeta_*^c$, $\eta_{\infty}^c$, $\theta_\infty^c$, and $\chi^c$ for various dimension in the range $d=3$ through $d=20$, with analytic results in $d=2$ and $d\rightarrow \infty.$}
\label{tab:numerics}
\end{table}

\subsection{Conformally Invariant Separation of Boundary Regions}
\label{sec:conformal}

Thus far, we have phrased our results entirely in terms of the $(\eta, \zeta)$ system of coordinates introduced in subsection \ref{sec:coordinates}. A coordinate-dependent characterization of the phase transition makes interpreting our results difficult; for example, while we showed in subsection \ref{sec:large-d-MI} that the boundary separation of antipodal caps at the phase transition scales like $1/d$ in the limit $d \rightarrow \infty$, this scaling could easily be changed by transforming to a $d$-dependent system of coordinates. It will be useful, instead, to formulate a coordinate-covariant notion of the separation between antipodal caps on the boundary of $\mathbb{H}^d$. The most obvious notion of boundary separation, the minimum proper distance between the surfaces, is ill-defined, since the boundary metric is only defined up to a conformal factor. Instead, we define a notion of distance between the caps that is both coordinate-independent and conformally invariant as follows.

In previous subsections, we considered the mutual information between equal-size antipodal caps on $S^{d-1}.$ To define a conformally invariant distance between caps, let us first generalize to consider antipodal caps $A$ and $B$ that are not necessarily the same size. For any such pair of caps, we define the \emph{conformal radii} as
\begin{equation}
	\gamma_{A,B} = \cosh(\eta_{A,B}) - \sinh(\eta_{A,B}),
\end{equation}
where $\eta_{A,B}$ is the value of $\eta$ at the cap boundary $\del A$ (equivalently, $\del B$). The conformal cross-ratio
\begin{equation} \label{eq:chi}
	\chi = \frac{(\gamma_A - \gamma_B)^2}{\gamma_A \gamma_B}
\end{equation}
can be shown to be invariant under conformal transformations. We show this explicitly in Appendix \ref{app:cross-ratio}, where we also show that $\gamma_A$ $(\gamma_B)$  is the radius of the sphere obtained by mapping $A$ ($B$) to Euclidean space conformally via the stereographic projection. This is the sense in which $\chi$ encodes a conformally invariant distance between the caps. When $A$ and $B$ are equal size, \eqref{eq:chi} simplifies to
\begin{equation}
	\chi = 4 \sinh^2(\eta_A),
\end{equation}
where we have used $\eta_B = - \eta_A.$

In addition to being invariant under passive conformal transformations that change the metric, $\chi$ also has the advantage of being invariant under \emph{active} conformal transformations that move the regions $A$ and $B$ around the sphere. Since any two non-adjacent caps on $S^d$ can be mapped to equal-size antipodal caps by a conformal transformation, any two caps develop nonzero mutual information at order $G_N^{-1}$ exactly when their cross-ratio $\chi$ equals the cross-ratio $\chi^c$ of equal-size caps at the phase transition. The appropriate generalization of $\chi$ for non-antipodal caps was introduced in \cite{Herzog:2014fra} and is reproduced in equation \eqref{eq:herzog-cross-ratio} of Appendix \ref{app:cross-ratio}. To check whether two ball-shaped regions have nonzero mutual information at leading order in $G_N$, therefore, one need only compute $\chi$ using equation \eqref{eq:herzog-cross-ratio} and compare it to the critical value $\chi^c$ given in Table \ref{tab:numerics} for the dimension in question.

By using equation \eqref{eq:eta-infty-star} for the value of $\eta$ at which the phase transition occurs in the limit $d\rightarrow \infty$, we see that in the limit of large $d$, the critical value $\chi^c$ is given by
\begin{equation} \label{eq:critical-chi-large-d}
	\chi^c = \frac{1}{(d-2)^2} \frac{\pi^2}{4} + O\left(\frac{1}{(d-2)^4} \right).
\end{equation}
Since $\chi$ represents a conformally invariant distance between the caps (cf. equation \eqref{eq:chi}), the fact that $\chi^c$ vanishes in the limit $d \rightarrow \infty$ indicates that equal size, antipodal caps must be arbitrarily close to one another to develop nonzero mutual information at order $1 / G_N$ in the large-$d$ limit.

\section{Monogamy of Entanglement in Many-Dimensional, Regular Lattices}
\label{sec:monogamy}

With the holographic calculation of Section \ref{sec:geometry} now completed, we turn to interpreting our results from the perspective of quantum information theory. In the introduction to this paper, we argued that the main technical result of Section \ref{sec:geometry}, which shows that antipodal boundary regions in the vacuum of a holographic theory must take up the entire volume of the boundary to develop nontrivial correlations at order $G_N^{-1}$, was one manifestation of a broader phenomenon of the spatial decoupling of entanglement in the large-$d$ limit. We also gave a heuristic argument based on monogamy of entanglement to explain this decoupling. The rough flavor of this argument was to suggest that in an isotropic state, each spatially local degree of freedom should be entangled with every spatial dimension equally; the principle of monogamy of entanglement should suggest, then, that it must be relatively unentangled with its neighbors in any given direction in the limit $d \rightarrow \infty.$ The ``monogamy of entanglement,'' however, is a general principle, not a single quantitative statement: it is realized in many different forms, mostly taking the form of various \emph{quantum de Finetti theorems} that restrict the entanglement structure of states obeying certain kinds of symmetry. The purpose of this section is to use these theorems to put our heuristic ``isotropy argument'' on concrete footing, and to argue rigorously that the spatial decoupling of ground state entanglement in the large-$d$ limit is a generic feature of spatially local, isotropic quantum systems.\footnote{\ 
Note that in Section \ref{sec:geometry}, we showed the large-$d$ decoupling of ground state correlations by computing the mutual information, which counts both quantum and classical correlations. The monogamy of entanglement arguments presented here explain only the generic decoupling of \emph{quantum} correlations; the large-$d$ decoupling of classical correlations found in Section \ref{sec:geometry} may be a special feature of holographic theories. We discuss this further at the end of this section.}

Before proceeding, we pause to review some general properties of entangled states. A pure state $\ket{\psi}_{AB}$ on a bipartite Hilbert space $\mathcal{H}_A \otimes \mathcal{H}_B$ is said to be unentangled if it can be written as a product state,
\begin{equation} \label{eq:unentangled-pure}
	\ket{\psi}_{AB} = \ket{\psi}_{A} \otimes \ket{\psi}_{B}.
\end{equation}
A general density matrix $\rho^{AB}$ --- i.e., a positive semidefinite Hermitian operator with unit trace --- is said to be unentangled if it can be written as a \emph{classical mixture of product states}:
\begin{equation} \label{eq:unentangled-mixed}
	\rho^{AB}
		= \sum_{j} p_j \rho_{j}^{A} \otimes \rho_j^{B}.
\end{equation}
Here $\rho_j^{A}$ and $\rho_{j}^B$ are density operators on systems $A$ and $B$, and $\{p_j\}$ is a set of probabilities satisfying $\sum_j p_j = 1.$ States of this form are often called \emph{separable}. States of the form \eqref{eq:unentangled-pure} and \eqref{eq:unentangled-mixed} are called unentangled because they can be prepared starting from any product state on $\mathcal{H}_A \otimes \mathcal{H}_B$ by performing local unitaries on the individual subsystems and exploiting classical randomness; no joint unitary operation is present to generate entanglement between the systems. To prepare a state of the form \eqref{eq:unentangled-pure}, one simply prepares the pure states in systems $A$ and $B$ separately; to prepare a state of the form \eqref{eq:unentangled-mixed}, one first draws a random number $j$ from the distribution $\{p_j\}$, then prepares the corresponding state $\rho_{j}^{A} \otimes \rho_j^{B}$ by acting on $\mathcal{H}_{A}$ and $\mathcal{H}_{B}$ with local unitaries. This procedure will produce, for any measurement, exactly the same statistics as would be measured in the state \eqref{eq:unentangled-mixed}. More abstractly, one can define the class of separable states \eqref{eq:unentangled-mixed} as the set of bipartite states that can be prepared from a product state using only \emph{local operations and classical communication} (LOCC); if it is possible to reproduce the measurement statistics of a state without ever performing a nonlocal operation to couple the systems, they cannot meaningfully be said to be entangled. Our goal in the following, as is common in the quantum information literature, will be to argue that large-$d$, isotropic ground states are locally very close to being separable; this is the sense in which their entanglement structure decouples at large $d$.\footnote{\ 
Familiar entanglement measures such as the von Neumann entropy are useful only in certain contexts; for example, the von Neumann entropy of a subsystem is only a good entanglement measure when the full state is pure. The distance from a given state to the nearest separable state is the only universal measure of entanglement in arbitrary settings.}

\begin{figure}[h]
\centering
\subfloat[\label{fig:square-lattice}]{
\includegraphics[scale=1]{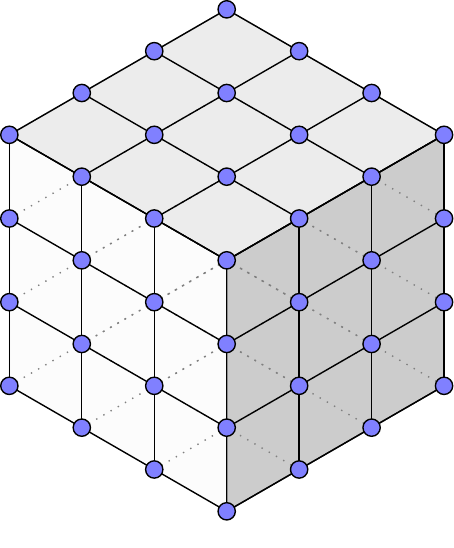}
}
\hspace{1.5cm}
\subfloat[\label{fig:square-state}]{
%
%
%
\includegraphics[scale=1]{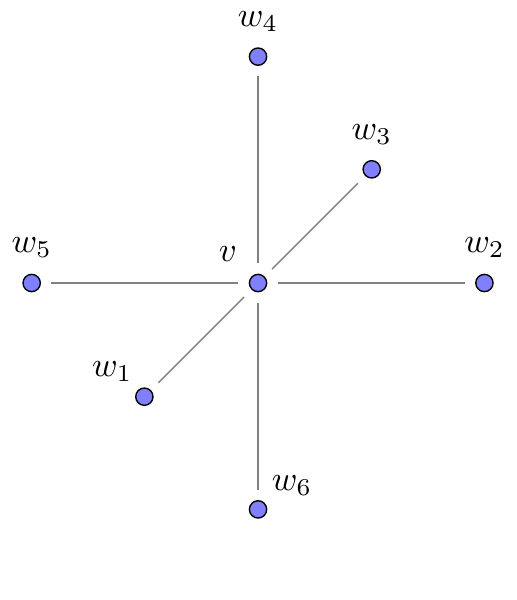}
}
\caption{(a) A square lattice in $d=3.$ Here each blue dot represents a local Hilbert space $\mathbb{C}^k.$ (b) A visual representation of the reduced ground state on a single vertex $v$ and its neighbors $\{w_1, \dots w_6\}.$}
\label{fig:square-pics}
\end{figure}

We take as our model a quantum system on a $d$-dimensional, regular lattice\footnote{\ 
Formally, a \emph{regular lattice} is one whose symmetry group acts transitively on the set of \emph{flags}, where a flag is a tuple consisting a vertex, an edge containing that vertex, a face containing that edge, and so on up to the highest-dimension object permitted by the lattice. In two dimensions, the square, triangular, and hexagonal lattices are all examples of regular lattices.} with finite local Hilbert space dimension $k$ and lattice symmetry group $G$. In practice, we will not actually use the fact that the lattice is regular, only that (i) the degree of each vertex (i.e., the number of neighbors) scales linearly with $d$, and (ii) for any fixed vertex $v$ and two neighboring vertices $w$ and $w'$, the symmetry group $G$ contains an element that maps $w$ to $w'$ while keeping $v$ fixed. Condition (ii) is that the graph is \emph{isotropic}. For concreteness, one can picture the generalized, $d$-dimensional square lattice with Hilbert space $\mathbb{C}^k$ on each vertex; this is sketched in Figure \ref{fig:square-pics} for the case $d=3.$ Our goal will be to show that in the large-$d$ limit, any two neighboring vertices in the ground state of such a system are unentangled up to corrections of order $O(1/d)$ in an appropriate norm.\footnote{\ 
The results on this section are quite similar in spirit to a well-known result in the quantum information literature, which is that the ground state of a maximally connected lattice is locally unentangled in the limit of infinitely many lattice sites \cite{mean-fields}. Generalizations of this result are known under the common mantra that ``mean field theory is exact in the limit of infinite degree.'' (See, e.g., \cite{mean-fields-1, mean-fields-2, mean-fields-3}.) What differs here is not so much our method as our perspective; we use the quantum de Finetti theorems not to show that highly-connected lattices are well-approximated by separable ground states, but to argue that the local decoupling of entanglement is a generic feature of large-$d$, isotropic systems.} We will see that this phenomenon follows directly from the isotropic symmetry of the graph, in direct analogy with the heuristic isotropy argument made in the introduction of this paper.

\begin{figure}[h]
\centering
\includegraphics[scale=1.05]{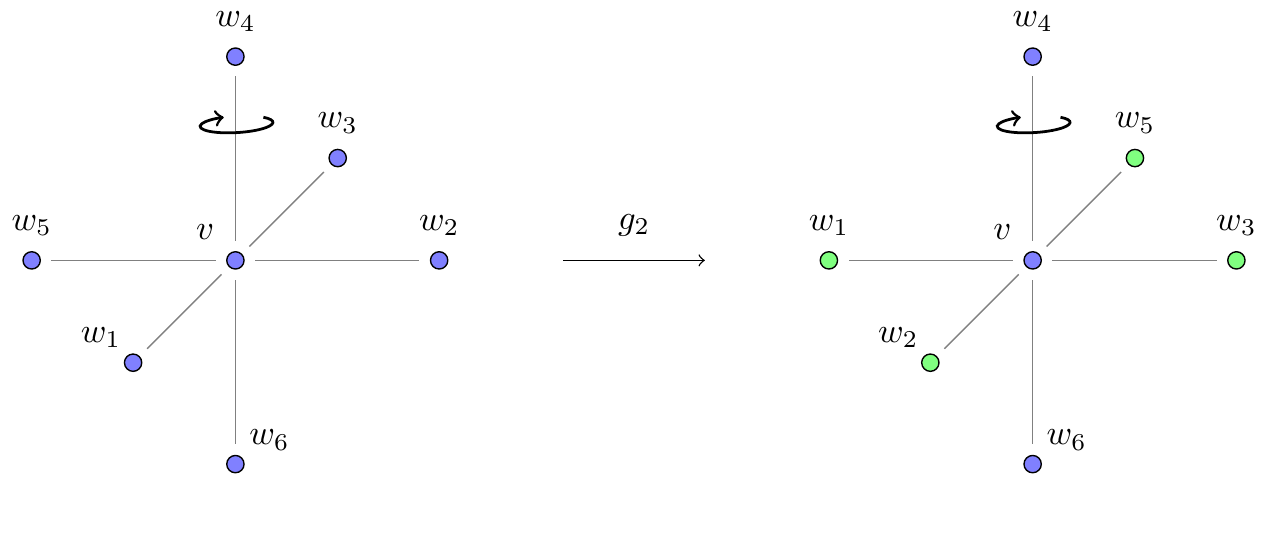}
\caption{The reduced state $\rho^{v w_1 \dots w_6}$ on the $3$-dimensional square lattice transforms under a rotation about the marked axis in a way that maps the label $w_1$ to the label $w_2$ while keeping $v$ fixed. Vertices that are altered by this symmetry transformation are marked in green in the second image. The existence of such a symmetry implies that the two-party reduced states satisfy $\rho^{v w_1} = \rho^{v w_2}.$ This rotation is just one choice of symmetry $g_2$; other lattice symmetries can also map $w_1$ to $w_2$ while differing in how they permute the other neighbors $w_i.$}
\label{fig:local-isotropy}
\end{figure}

Suppose we endow our lattice Hilbert space with a Hamiltonian $H$ that (i) commutes with each element of $G$, and (ii) has a unique ground state $\ket{0}.$ For any vertex $v$ with neighbors $\{w_1, \dots, w_n\}$, the reduced density matrix
\begin{equation}
	\rho^{v w_1 \dots w_n} = \tr_{\{v, w_1, \dots, w_n\}^c} \ket{0} \bra{0}
\end{equation}
describes the ground state of $v$ and its nearest neighbors. This state is sketched in Figure \ref{fig:square-state}. Since $H$ commutes with each element of $G$ and has a unique ground state, $\ket{0}$ is symmetric with respect to the lattice symmetries; this symmetry is inherited by the reduced state on $\{v, w_1, \dots, w_n\}$, so that $\rho^{v w_1 \dots w_n}$ satisfies
\begin{equation} \label{eq:lattice-symmetry}
	g \rho^{v w_1 \dots w_n} g^{-1}
		= \rho^{v w_1 \dots w_n}
\end{equation}
for any element $g \in G$. By our isotropy assumption, $G$ contains at least one element $g_j$ that maps $w_1$ to $w_j$ without moving $v$ for any $j \in \{1 \dots n\}.$ Applying this symmetry to equation \eqref{eq:lattice-symmetry} yields the expression
\begin{equation} \label{eq:permute-DM}
	\rho^{v w_1 \dots w_n}
		= g_j \rho^{v w_1 \dots w_n} g_j^{-1}
		= \rho^{v w_j \dots},
\end{equation}
where the ordering of the unlabeled systems on the right-hand side of this expression depends on the particular choice of $g_j.$  In the square lattice sketched in Figure \ref{fig:square-pics}, such a symmetry can be implemented, for example, by a rotation about $v$; in Figure \ref{fig:local-isotropy}, we sketch the action of one such symmetry transformation $g_2$ for the $3$-dimensional square lattice. By tracing out all but the first two subsystems in equation \eqref{eq:permute-DM}, we find that the biparite reduced density matrices on pairs of sites $\{v, w_j\}$ satisfy
\begin{equation} \label{eq:n-extension}
	\rho^{v w_1}
		= \rho^{v w_j}.
\end{equation}
In other words, \emph{any two nearest-neighbor density matrices sharing a vertex are identical}.

Intuitively, the principle of monogamy of entanglement suggests that the bipartite entanglement between $v$ and any of its neighbors $w_j$ must be small when the total number of neighbors $n$ is large; in order for $v$ to be entangled in the same way with a large number of neighbors, it cannot possibly be highly entangled with any of them. The quantitative version of this statement is as follows: a state $\rho^{v w_1 \dots w_n}$ satisfying equation \eqref{eq:n-extension} is called an \emph{n-extension} for the state $\rho^{v w_1}$; a bipartite state admitting an $n$-extension is unentangled in the sense of equation \eqref{eq:unentangled-mixed} up to $O(1/n)$ corrections. The proof of this claim follows from the quantum de Finetti theorems for symmetric states, which we will now introduce. The general idea of the following is to exploit the lattice symmetry, which guarantees the validity of equation \eqref{eq:n-extension} and hence implies that each vertex is equally entangled with all of its neighbors, to show that each two-party reduced state must be nearly separable.

The quantum de Finetti theorems govern how close a quantum state respecting a particular symmetry is to being unentangled. The quantum de Finetti theorem for symmetric states, first proved in \cite{qdF-1, qdF-2, qdF-3} and generalized to the form we use here by \cite{big-qdF}, states that a pure state $\ket{\Phi} \in (\mathbb{C}^{k})^{\otimes n}$ that is invariant under permutations of the $n$ subsystems is close to being separable on any $m$ subsystems with $m k \ll n.$ Formally, for any $m < n,$ there exists a probability measure $\dd m(\sigma)$ on the space of density matrices over $\mathbb{C}^k$ such that the $m$-party reduced state $\rho^{m}$ satisfies \cite{big-qdF}
\begin{equation} \label{eq:de-finetti-pure}
	\lVert \rho^m - \int \dd m(\sigma)\, \sigma^{\otimes m} \rVert_1 \leq 4 \frac{m k}{n}.
\end{equation}
In other words, the reduced state $\rho^m$ is within $O(mk/n)$ distance of the separable state $\int \dd m(\sigma) \sigma^{\otimes m}$. The norm that appears in this equation is the \emph{trace norm}, given by $\lVert A \rVert_1 = \tr\sqrt{A^{\dagger} A}.$ Density matrices that are close in the trace norm have similar expectation values on bounded operators; they satisfy
\begin{equation} \label{eq:holder}
	|\tr(\rho \mathcal{O}) - \tr(\sigma \mathcal{O})|
		\leq \lVert \rho - \sigma \rVert_1 \sup_{\ket{\psi} \in \mathcal{H}} \bra{\psi}{\mathcal{O}}\ket{\psi}
\end{equation}
for any operator $\mathcal{O}$. When the local Hilbert space dimension $k$ and the number of subsystems $m$ are much smaller than the total number of systems $n$, \eqref{eq:de-finetti-pure} implies that the $m$-partite reduced state of a symmetric state $\ket{\Phi}$ will reproduce, for any bounded operator, measurement statistics very close to that of an unentangled state.

A generalization of equation \eqref{eq:de-finetti-pure} for mixed states, also proved in \cite{big-qdF}, states that when an $n$-party density matrix $\rho^n$ commutes with permutations of the $n$ subsystems, the $m$-party reduced state satisfies
\begin{equation} \label{eq:de-finetti-mixed}
	\lVert \rho^m - \int \dd m(\sigma)\, \sigma^{\otimes m} \rVert_1 \leq 4 \frac{m k^2}{n}
\end{equation}
for some probability measure $\dd m(\sigma).$ The only difference from equation \eqref{eq:de-finetti-pure} is an extra factor of the local Hilbert space dimension $k$. A further generalization from \cite{big-qdF}, and the one that we will apply to our lattice system, concerns states that are symmetric \emph{relative to an auxiliary system}. It states that when a state $\rho^{A n}$ on the Hilbert space $\mathcal{H}_{A} \otimes (\mathbb{C}^k)^{\otimes n}$ commutes with permutations of the $n$ subsystems, the reduced state $\rho^{A m}$ for $m < n$ satisfies
\begin{equation} \label{eq:de-finetti-relative}
	\lVert \rho^{A m} - \int \dd m(\sigma)\, \tau_\sigma^{A} \otimes \sigma^{\otimes m} \rVert_1 \leq 4 \frac{m k^2}{n},
\end{equation}
where $\tau_\sigma^A$ is a $\sigma$-dependent family of density matrices on $\mathcal{H}_{A}.$ Put simply, equation \eqref{eq:de-finetti-relative} states that any system $\mathcal{H}_{A}$ that couples to a large number of identical systems in a permutation-invariant way must not be very entangled with any of them; its joint reduced state with any small collection of subsystems $m$ satisfying $m k^2 \ll n$ is close to separable.

We argued above that in the unique ground state of a $d$-dimensional lattice system with local Hilbert space dimension $k$, the reduced state of a vertex $v$ with any one of its neighbors $w_i$ is \emph{$n$-extendible}, where $n$ is the degree of the vertex. More precisely, the reduced state $\rho^{v w_i}$ has an $n$-extension $\rho^{v w_1 \dots w_n}$ satisfying
\begin{equation} \label{eq:n-extension-2}
	\rho^{v w_i} = \rho^{v w_j}
\end{equation}
for any $j \in \{1, \dots, n\}.$ One cannot apply equation \eqref{eq:de-finetti-relative} directly to this state; the $n$-extension $\rho^{v w_1 \dots w_n}$ is not symmetric under permutations of the $n$ neighboring vertices, since the lattice symmetry does not induce a full permutation symmetry on the neighbors of $v$. If, however, one symmetrizes the state $\rho^{v w_1 \dots w_n}$ with respect to subsystems $\{w_1, \dots, w_n\}$ to obtain a symmetric state $\tilde{\rho}^{v w_1 \dots w_n}$, the entanglement structure of this symmetrized state can be constrained by equation \eqref{eq:de-finetti-relative}. Since the symmetrized and unsymmetrized states agree on two-party subsystems, i.e. $\rho^{v w_j} = \tilde{\rho}^{v w_j}$, the relative quantum de Finetti theorem \eqref{eq:de-finetti-relative} can then be used to constrain $\rho^{v w_j}$. These states must satisfy
\begin{equation}
	\lVert \rho^{v w_j} - \int \dd m(\sigma)\, \tau_{\sigma}^{v} \otimes \sigma \rVert_1 \leq \frac{8 k^2}{n}
\end{equation}
for some probability measure $\dd m(\sigma)$ over states on $w_i.$ Under our assumption that the degree of each vertex scales linearly with dimension, $n = \alpha d$, we find that each bipartite state of neighboring vertices $\rho^{v w_i}$ satisfies an equation of the form
\begin{equation} \label{eq:bipartite-decoupling}
	\lVert \rho^{v w_j} - \int \dd m(\sigma)\, \tau_{\sigma}^{v} \otimes \sigma \rVert_1 \leq \frac{8 k^2}{\alpha d},
\end{equation}
i.e., it is within $O(k^2/d)$ trace distance of being separable.

The arguments presented thus far in this section show that the spatial decoupling of entanglement in the large-$d$ limit is a generic feature of isotropic lattice systems with (i) finite local Hilbert space dimension, and (ii) a unique ground state. It follows that any quantum system that is well approximated by an isotropic lattice satisfying (i) and (ii) must also exhibit entanglement decoupling at large $d$. There are two main obstructions to applying this analysis toward understanding the decoupling of holographic correlations observed in Section \ref{sec:geometry}. The first is that the boundary theory in a holographic system is a Lorentzian quantum field theory, which naively has infinite local Hilbert space dimension even under an arbitrary lattice regularization. The second is that the decoupling of the mutual information found in Section \ref{sec:geometry} suggests a decoupling of both quantum \emph{and} classical correlations, while the de Finetti arguments presented in this section apply only to entanglement, i.e., to quantum correlations.

To address the first obstruction, we note that the ground state of any quantum field theory can be well approximated by a lattice-regulated field theory, which in turn can be well approximated at least in the ground state by keeping only a finite number of basis states on each vertex. The only question is how the local Hilbert space dimension $k$ required to obtain a good approximation scales with the spatial dimension $d$. Since equation \eqref{eq:bipartite-decoupling} shows that any two-party reduced state of an isotropic ground state is separable up to corrections of order $O(k^2/d)$, the quantum de Finetti theorem should suggest decoupling of entanglement in quantum field theories provided that the local Hilbert space dimension required to approximate the true ground state grows more slowly than $\sqrt{d}.$ At present, we have no general argument that this should be the case; however, we note that this condition can be relaxed if one uses a coarser notion of the ``distance to separability'' than distance in the trace norm. The trace distance $\lVert \rho^{AB} - \sigma^{AB} \rVert_1$ gives the maximum probability of distinguishing two states by performing a joint measurement on both systems. If we ask instead for the maximum probability of distinguishing the states by performing \emph{local} measurements on each subsystem, we are led to the \emph{LOCC norm} $\lVert \cdot \rVert_{\mathrm{LOCC}}$; it was shown in \cite{LOCC-decoupling} that an $n$-extendible state $\rho^{AB}$ is within $O(\sqrt{\log{k}/n})$ of being separable with respect to this norm. This norm measures how well two local observers sitting on neighboring sites would be able to decide whether or not their joint state is entangled; in order for the ground state of a quantum field theory to be locally decoupled in an \emph{operational} sense, therefore, we need only require that the local site dimension $k$ grows more slowly than $e^{d}.$

As for the second obstruction, we view this not as a true obstruction but as a feature of our calculation! The arguments of this section suggest that spatial decoupling of ground state entanglement is generic at large $d$; the calculations of Section \ref{sec:geometry}, however, suggest the decoupling of \emph{all} correlations at order $G_N^{-1}$, both quantum and classical. While the decoupling of entanglement is a natural consequence of isotropy combined with the monogamy of entanglement, the decoupling of classical correlations are more subtle. It is possible that the large-$d$ decoupling of classical correlations is in fact generic for reasons that are beyond the scope of this section, or it is possible that the large-$d$ decoupling of \emph{holographic} classical correlations is a genuinely special feature of holography.

Holographic systems are well known to have universal entanglement structure that is not generic in arbitrary quantum systems. For example, states with entropies obeying the HRRT formula \eqref{eq:HRT} are known to satisfy an entropy inequality known as the \emph{monogamy of mutual information} \cite{MMI, maximin}
\begin{equation}
	S(A) + S(B) + S(C) - S(AB) - S(BC) - S(AC) + S(ABC) \leq 0.
\end{equation}
In \cite{MMI}, this inequality was interpreted as implying that at order $G_N^{-1}$, quantum correlations dominate over classical ones in holographic systems. Our observation that all holographic correlations decouple at large $d$ is certainly compatible with this interpretation: quantum correlations decouple due to monogamy of entanglement, and it is certainly possible that classical correlations then decouple solely due to their subdominance. However, this argument, even if valid, does not explain the \emph{mechanism} by which the classical correlations decouple; understanding this mechanism, and understanding what it can teach us about the correlation structure of holographic quantum theories, remains an open question for future work.

\section{Discussion}
\label{sec:discussion}

The primary motivation for our analysis was to understand how the spatial correlation of spacetime regions in a quantum field theory depends on the dimension of the spacetime. To address this question, we focused our attention on holographic field theories where the difficult computation of the entanglement entropy of a given region in the boundary CFT$_{d}$ can be mapped to the simpler problem of computing the area of the smallest-area extremal surface in AdS$_{d+1}$ anchored to the boundary of the region via the HRRT formula \eqref{eq:HRT}. In this context, we analyzed the phase transition in the leading $1/G_{N}$ piece of the mutual information $I(A:B)$ between two equal-sized antipodal caps $A$ and $B$ in the vacuum state at a fixed time on the boundary of global AdS$_{d+1}$, whereby the HRT surface switches from two disconnected surfaces anchored on the boundary of each region to one connected surface anchored on the boundary of both regions.

Surprisingly, we found in Section \ref{sec:geometry} that in the limit of infinitely many spacetime dimensions, the regions $A$ and $B$ have to occupy the entire boundary volume in order to develop non-zero mutual information at order $G_N^{-1}$. This conclusion contradicts a naive argument presented in the Introduction, which used monogamy of entanglement to argue that some $O(1)$ fraction of the boundary degrees of freedom should suffice to develop correlations at order $G_N^{-1}$. This argument, however, depended on the assumption that ball-shaped regions are highly entangled with their complements at large $d$, whereas our calculations in Section \ref{sec:geometry} show that holographic correlations between spatial regions decouple in the limit of large $d$. 

Our analysis of this phase transition for the extremal surfaces in AdS$_{d+1}$ anchored on the boundary of $A$ and $B$ was greatly simplified by the use of a particular choice of coordinates adapted to an isometry of $\mathbb{H}^{d}$ that is generated by a coordinate Killing vector field. The advantage of these coordinates was that the disconnected HRT surfaces were level sets of the Killing coordinate, and the Euler-Lagrange equations for the connected, ``tube-like'' surfaces became first-order ODEs. This simplification enabled us to solve for the phase transition numerically in any spacetime dimension $d$ and analytically in the limit $d \to \infty$.


The mutual information is indeed but one of many diagnostics of the correlations between the degrees of freedom in a quantum field theory. However, we argued that the decoupling of the holographic mutual information in the limit of infinite spacetime dimensions is actually a manifestation of a much more general phenomenon of decoupling in quantum systems with a large number of spacetime dimensions. We demonstrated in Section \ref{sec:monogamy} that in the simple setting of a quantum system on a regular $d$-dimensional lattice, monogamy of entanglement implies that the entanglement between neighboring lattice sites in an isotropic state vanishes in the limit $d \to \infty$. More precisely, using the quantum de Finetti theorems, we showed that the reduced density matrix for any two neighboring vertices is within $1/d$ trace distance of a separable state. The primary technical result of Section \ref{sec:geometry}, that the mutual information in a holographic theory decouples spatially at large $d$, implies a stronger result than the monogamy of entanglement argument presented in Section \ref{sec:monogamy}: since mutual information counts both classical and quantum correlations, the decoupling of holographic mutual information at large $d$ is qualitatively more severe than the generic decoupling of quantum correlations implied by the quantum de Finetti theorems. A more precise interpretation of this result remains open for future work.

Our work inspires many other possible directions for future research.  A major success of the large-$d$ expansion in general relativity initiated in \cite{largeD}  is the remarkable agreement of the $1/d$ expansion, truncated at only a few orders, with numerical results at small $d$ for a wide range of physically interesting quantities (such as the quasinormal mode spectrum of a low-dimensional black brane, for example \cite{largeD}). One might expect that a similar agreement could be found between our numerical values for the phase transition at small $d$  (cf. Section \ref{sec:numerics}) and our $1/d$ expansion for the phase transition parameter (cf. Section \ref{sec:large-d-MI}). In fact, the result obtained for the phase transition parameter $\zeta_*^c$ by expanding to one further order in the $1/d$ expansion of equation \eqref{eq:integral-sum} differs from the actual numerical values computed in Section \ref{sec:numerics} by a fairly small percentage error: while the percentage error for the $d=3$ phase transition parameter is ${\sim}29\%$,  the error drops to ${\sim}3.5\%$ by $d=5$ and below $1\%$ by $d=18.$ By adding only a few more orders in the $1/d$ expansion, it should be possible to get good analytic approximations to the numerically computed phase transition parameters in low $d$. This suggests broader applicability of the large-$d$ program in holography: there are other interesting geometric quantities related to holographic entanglement entropy that are analytically formidable in physically interesting numbers of dimensions 
 that may yield to a truncated $1/d$ analysis.

In the Introduction we motivated our choice of the state by arguing that the vacuum state is the most difficult to unentangle so that the vacuum state has the best hope of remaining entangled at large $d$. It would nevertheless be interesting to generalize our analysis to other geometric states, such as perturbations to the vacuum state, which correspond to linear perturbations of pure AdS \cite{Faulkner:2013ica}. One particularly exciting prospect would be to analyze the mutual information structure of thermal states, i.e. black holes, at large $d$, either for spatial partitions of a single asymptotic region or for the mutual information between two asymptotic regions.\footnote{For some existing work in this direction, see \cite{other-spacetimes-1, other-spacetimes-2, other-spacetimes-3, other-spacetimes-4}.} In the second case, phase transitions in the mutual information between two asymptotic regions in certain black hole spacetimes have been shown to be related \cite{Shenker:2013pqa} to the phenomenon of scrambling; it would be interesting to investigate such phenomena in the large-$d$ limit.

One important open question in our whole discussion of the mutual information phase transition is to understand directly from the CFT side of the duality why the regions $A$ and $B$ need to collectively occupy the entire boundary volume to develop non-zero mutual information at order $G_N^{-1}$ in the limit $d \to \infty$. This problem is in principle tractable due to the simplification of conformal blocks at large $d$ \cite{large-d-blocks}. However, our attempts to compute the mutual information directly in the CFT have been unsuccessful due to issues in computing $S(A \cup B)$ with the replica trick. To compute the joint Renyi entropy $S^{(n)}(A \cup B) = (1-n)^{-1}\log\mathrm{Tr}(\rho^{n}) = (1-n)^{-1}\log\langle\tau_{A \cup B}\rangle$, where $\tau_{A \cup B}$ is a twist operator for $n$ copies of the CFT whose support is the entangling surface $\partial A \cup \partial B$, one needs to evaluate the one-point function of a disconnected, non-local operator in $n$ copies of the CFT. This seems to be equally difficult in any spacetime dimension $d > 2$ and does not seem to simplify at large $d$ or large $N$. In the limit $\chi \to \infty$\footnote{\ 
Recall equation \eqref{eq:chi} for the definition of the conformal invariant $\chi$.}, it has been argued that $\langle\tau_{A \cup B}\rangle$ can be written as a sum of products of local operators in each copy of the CFT inserted at points in $A$ and $B$ so that the problem becomes doable \cite{Cardy:2013nua}. But according to equation \eqref{eq:critical-chi-large-d}, this is the opposite limit from the one where the phase transition occurs! If one could reproduce our result for the mutual information phase transition directly from the CFT, it would not only provide a nice test of the HRRT formula, but it would also give further intuition as to why spatial regions decouple in the large $d$ limit.

Let us conclude with some more general prospects for future directions. While the idea of large $d$ is not new in quantum gravity \cite{Strominger:1981jg}, to our knowledge this work is the first attempt to employ the large-$d$ limit in the context of quantum information in holography. Given the success of our approach in obtaining an analytic formula for the phase transition in the leading $G_{N}$ term in the mutual information, it is natural to ask whether the large $d$ limit could be applied to other problems in the AdS/CFT correspondence that have hitherto been intractable in a low number of dimensions. Are there other holographic entanglement phenomena that similarly simplify in the large-$d$ limit? A basic question that one would like to answer is what are the necessary and sufficient criteria for a CFT state to have a nice dual geometry. 
A recent research avenue approaches the question by  exploring the restrictions on the entanglement structure due to entropy inequalities.  The program initiated in  \cite{Hubeny:2018ijt,Hubeny:2018trv} obtains these via an algebraic method which has a close connection to the marginal independence problem (namely when can a given set of mutual informations between composite subsystems simultaneously vanish) \cite{CHRR}.  The large-$d$ decorrelation indicates that in this limit, holographic configurations are maximally localized in entropy space.

From the bulk side, the geometric toolkit may become more tractable at large $d$. 
As a concrete calculation, one could study whether shape deformations of the entanglement entropy simplify in the large-$d$ limit. In particular, a naive large-$d$ intuition might suggest that the way a given extremal surface corresponds to the shape of the entangling surface is much more ``local'' in higher dimensions, so that a given bulk region can be more readily decoded from a more limited set of specific entanglement entropies.  Much more ambitiously, one could even try to construct the bulk Einstein's equations beyond the linear approximation from entanglement entropy relations \cite{Faulkner:2013ica,Swingle:2014uza}. Pushing this one step further, perhaps the very machinery of bulk reconstruction simplifies in the large-$d$ limit so that we can really see the inner workings of the bulk-boundary duality in action. 

\acknowledgments{We acknowledge Aidan Chatwin-Davies, Roberto Emparan, Marina Mart\'{i}nez, Onkar Parrikar, Geoff Penington, Dan Ranard, and Mukund Rangamani for useful conversations. JS thanks the Center for Quantum Mathematics and Physics at UC Davis for their hospitality when this work was initiated. Portions of this work were completed at the Yukawa Institute for Theoretical Physics during workshop YITP-T-19-03 and at the Centro de Ciencias de Benasque during the 2019 workshop ``Gravity: new perspectives from strings and higher dimensions.'' SC-E and VH are supported by U.S.\ Department of Energy grant DE-SC0009999 and by funds from the University of
California. BEN is an FWO [PEGASUS]$^2$ Marie Sk\l{}odowska-Curie Fellow supported by the FWO and European Union's Horizon 2020 research and innovation program under the Marie Sk\l{}odowska-Curie grant agreement No.\ 665501, and by ERC grant ERC-2013-CoG 61673 HoloQosmos. JS is supported by AFOSR (FA9550-16-1-0082), the Simons It from Qubit collaboration, and DOE Award No.\ DE-SC0019380.}

\appendix

\section{Conformal Cross-Ratio for Antipodal Caps}
\label{app:cross-ratio}

In this Appendix, we demonstrate that the conformal cross-ratio $\chi$ defined by \eqref{eq:chi} is the unique conformal invariant characterizing the configuration of two antipodal caps on $S^{d-1}$. The conformal cross-ratio $\chi$ can be thought of as the conformally invariant distance between the caps.

\begin{figure}[h]
\centering
\includegraphics{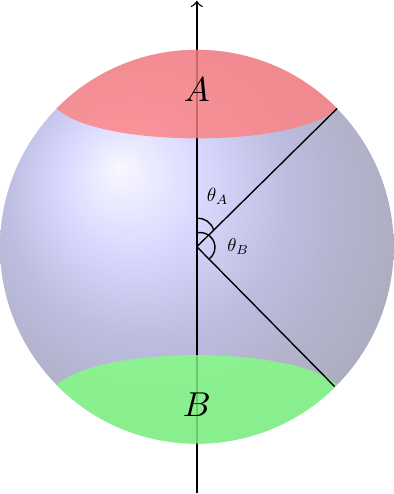}
\caption{Two polar cap regions $A$ and $B$ centered around antipodal points of $S^{2}$.}
\label{fig:polarcaps_cyl}
\end{figure}

Consider the boundary cylinder $\mathbb{R} \times S^{d-1}$ of $\mathrm{AdS}_{d+1}$ in the coordinates $(t,\theta,\Omega_{d-2})$ with metric given by
\begin{equation}
	ds^2 = -dt^2+d\theta^{2}+\sin^{2}\theta\,d\Omega_{d-2}^2.
\end{equation}
Let $A$ and $B$ be polar caps on the any constant-$t$ slice centered at $\theta=0$ and $\theta = \pi$, respectively. Since the metric on the cylinder is time translation-invariant, we take $t=0$ in the following discussion. Each polar cap is defined by the inclination angle $\theta$ of its boundary:
\begin{eqnarray}
	A 
		& = & \{(\theta,\Omega_{d-2}) \in S^{d-1}\,|\,\theta \leq \theta_{A}\}, \\
	B
		& =&  \{(\theta,\Omega_{d-2}) \in S^{d-1}\,|\,\theta \geq \theta_{B}\},
\end{eqnarray}
with $\theta_A \leq \theta_B$, as illustrated in Figure \ref{fig:polarcaps_cyl}.

\begin{figure}[h]
\centering
\includegraphics{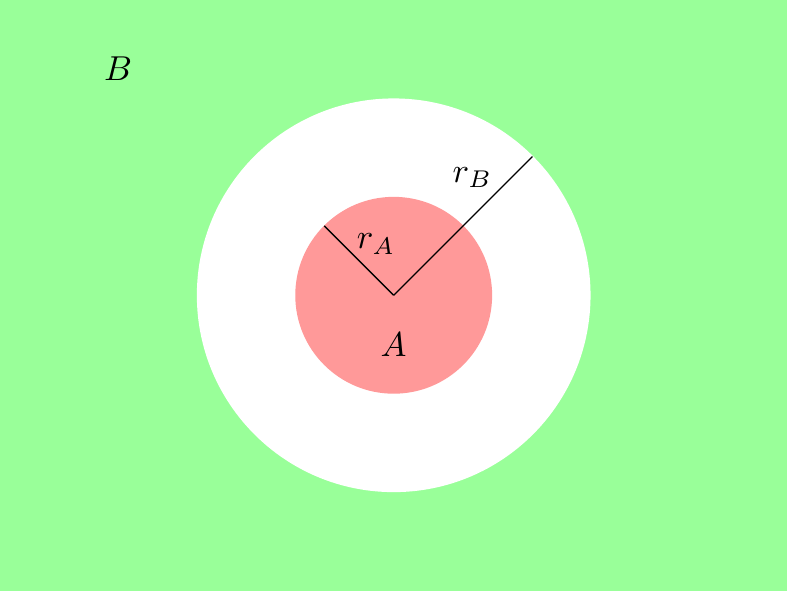}
\caption{The polar caps $A$ and $B$ on the $t=0$ slice of $\mathbb{R} \times S^{2}$ are mapped to balls centered at $r=0$ and $r=\infty$ on the $\tau=0$ slice of $\mathbb{R}^{1,2}$.}
\label{fig:polarcaps_Mink}
\end{figure}

To determine the conformally invariant distance between the caps $A$ and $B$, we first conformally map the cylinder $\mathbb{R} \times S^{d-1}$ to Minkowski spacetime $\mathbb{R}^{1,d-1},$ where the action of the conformal group is simpler. Writing $\mathbb{R}^{1,d-1}$ in spherical coordinates $(\tau,r,\Omega_{d-2})$, we choose a conformal map given by the stereographic projection
\begin{equation}
	\tau \pm r = \tan\bigg(\frac{t\pm\theta}{2}\bigg).
\end{equation}
Observe that the constant $t=0$ slice on the cylinder is mapped to the constant $\tau = 0$ slice in Minkowski spacetime. The polar caps $A$ and $B$ are mapped to $(d-1)$-dimensional spatial balls centered at $r=0$ and $r=\infty$, respectively. Explicitly,
\begin{equation}
	A \mapsto \mathcal{B}_{r_{A}}(r=0), \qquad r_{A} = \tan\bigg(\frac{\theta_{A}}{2}\bigg)
\end{equation}
and
\begin{equation}
	B \mapsto \big(\mathcal{B}_{r_{B}}(r=0)\big)^{c}, \qquad r_{B} = \tan\bigg(\frac{\theta_{B}}{2}\bigg),
\end{equation}
as illustrated in Figure \ref{fig:polarcaps_Mink}.

This map makes it clear that the angular distance $\theta_{B}-\theta_{A}$ between the caps is not invariant under dilatations $r_{A,B} \mapsto \lambda r_{A,B}$ ($\lambda \in \mathbb{R}$). To construct a conformally invariant quantity $\chi$ that characterizes this configuration of balls in Minkowski spacetime, we first observe that the only translation- and rotation-invariant quantities describing a configuration of two spheres are the radii $r_{A,B}$ of the spheres and the distance $\ell$ between their centers. (While $\ell$ vanishes for the configuration of spheres we are currently considering, special conformal transformations will make it nonzero.) Up to taking an overall power, the only functions $\chi(r_A, r_B, \ell)$ that are invariant under dilatations are ratios of homogeneous polynomials of equal degree, i.e.,
\begin{equation} \label{eq:chi-general}
	\chi(r_A, r_B, \ell)
		= \frac{\sum_{m+n+p=k} a_{m,n,p} r_A^{m} r_B^{n} \ell^p}{\sum_{m+n+p=k} b_{m,n,p} r_A^m r_B^n \ell^p}.
\end{equation}
Under a special conformal transformation
\begin{equation}
	x^{\mu} \mapsto \frac{x^{\mu} - b^{\mu} x^2}{1 - 2 b \cdot x + b^2 x^2},
\end{equation}
a configuration of two spheres centered around the origin transforms to a new configuration of spheres according to
\begin{eqnarray}
	r_{A,B} & \mapsto & r_{A',B'} = \frac{r_{A, B}}{|1 - b^2 r_{A, B}^2|}, \\
	\ell^2 = 0  & \mapsto & \ell^2 = \frac{b^2 (r_A^2 - r_B^2)^2}{(1 - b^2 r_A^2)^2 (1 - b^2 r_B^2)^2}.
\end{eqnarray}
Imposing invariance of equation \eqref{eq:chi-general} under transformations of this form uniquely fixes $\chi$ to take the form
\begin{equation}\label{eq:chi_Mink_pre}
\chi = \frac{(r_{A}-r_{B}-\ell)(r_{A}-r_{B}+\ell)}{r_{A}r_{B}}
\end{equation}
up to multiplication by a constant and taking the reciprocal. For concentric spheres --- i.e., antipodal caps on $\mathbb{H}_d$ --- this reduces to
\begin{equation}\label{eq:chi_Mink}
\chi = \frac{(r_{A}-r_{B})^{2}}{r_{A}r_{B}}
\end{equation}
Using \eqref{eq:theta-transform}, it is straightforward to show that the ball radii $r_{A,B}$ are equivalent to the conformal radii $\gamma_{A,B}$ introduced in Section \ref{sec:conformal}. We thus recover the form of $\chi$ given in the main text by equation \eqref{eq:chi}.

An equivalent construction of $\chi$ due to \cite{Herzog:2014fra} is as follows: consider a line in $\mathbb{R}^{d-1}$ that passes through the centers of both balls. The intersection of this line with the boundary of each ball defines four points: $p_1$ and $p_2$ for the first ball, and $q_1$ and $q_2$ for the second. The conformal cross-ratio
\begin{equation} \label{eq:herzog-cross-ratio}
	\chi = 4 \frac{|p_1 - q_1| |p_2 - q_2|}{|p_1 - p_2| |q_1 - q_2|}
\end{equation}
is then a conformal invariant that reproduces \eqref{eq:chi_Mink_pre}. To find the value of $\chi$ for a particular configuration of caps on $S^{d-1}$, one simply maps the two caps to $\mathbb{R}^{d-1}$ conformally, draws a line through the centers of the corresponding balls, and constructs the quantity \eqref{eq:herzog-cross-ratio}. This construction illustrates that the space of configurations of two non-overlapping balls in $\mathbb{R}^{d-1}$ up to translations and rotations is actually one-dimensional; this is why there is only one conformal invariant up to powers and scalar multiplication.

\bibliographystyle{JHEP}
\bibliography{phase-transition}

\providecommand{\href}[2]{#2}\begingroup\raggedright\begin{thebibliography}{10}

\bibitem{Witten:2018lha}
E.~Witten, \emph{{APS Medal for Exceptional Achievement in Research: Invited
  article on entanglement properties of quantum field theory}},
  \href{https://doi.org/10.1103/RevModPhys.90.045003}{\emph{Rev. Mod. Phys.}
  {\bfseries 90} (2018) 045003}
  [\href{https://arxiv.org/abs/1803.04993}{{\ttfamily 1803.04993}}].

\bibitem{RT}
S.~Ryu and T.~Takayanagi, \emph{Aspects of holographic entanglement entropy},
  {\emph{JHEP} {\bfseries 2006} (2006) 045}
  [\href{https://arxiv.org/abs/hep-th/0605073}{{\ttfamily hep-th/0605073}}].

\bibitem{HRT}
V.~E. Hubeny, M.~Rangamani and T.~Takayanagi, \emph{A covariant holographic
  entanglement entropy proposal}, {\emph{JHEP} {\bfseries 2007} (2007) 062}
  [\href{https://arxiv.org/abs/0705.0016}{{\ttfamily 0705.0016}}].

\bibitem{LM}
A.~Lewkowycz and J.~Maldacena, \emph{Generalized gravitational entropy},
  {\emph{JHEP} {\bfseries 2013} (2013) 90}
  [\href{https://arxiv.org/abs/1304.4926}{{\ttfamily 1304.4926}}].

\bibitem{DLR}
X.~Dong, A.~Lewkowycz and M.~Rangamani, \emph{Deriving covariant holographic
  entanglement}, {\emph{JHEP} {\bfseries 2016} (2016) 28}
  [\href{https://arxiv.org/abs/1607.07506}{{\ttfamily 1607.07506}}].

\bibitem{FLM}
T.~Faulkner, A.~Lewkowycz and J.~Maldacena, \emph{Quantum corrections to
  holographic entanglement entropy}, {\emph{JHEP} {\bfseries 2013} (2013) }
  [\href{https://arxiv.org/abs/1307.2892}{{\ttfamily 1307.2892}}].

\bibitem{Rangamani:2016dms}
M.~Rangamani and T.~Takayanagi, \emph{{Holographic Entanglement Entropy}},
  \href{https://doi.org/10.1007/978-3-319-52573-0}{\emph{Lect. Notes Phys.}
  {\bfseries 931} (2017) pp.1}
  [\href{https://arxiv.org/abs/1609.01287}{{\ttfamily 1609.01287}}].

\bibitem{cutoffs}
J.~Sorce, \emph{Holographic entanglement entropy is cutoff-covariant},
  \href{https://arxiv.org/abs/1908.02297}{{\ttfamily 1908.02297}}.

\bibitem{Wolf:2007tdq}
M.~M. Wolf, F.~Verstraete, M.~B. Hastings and J.~I. Cirac, \emph{{Area Laws in
  Quantum Systems: Mutual Information and Correlations}},
  \href{https://doi.org/10.1103/PhysRevLett.100.070502}{\emph{Phys. Rev. Lett.}
  {\bfseries 100} (2008) 070502}
  [\href{https://arxiv.org/abs/0704.3906}{{\ttfamily 0704.3906}}].

\bibitem{Cardy:2013nua}
J.~Cardy, \emph{{Some results on the mutual information of disjoint regions in
  higher dimensions}},
  \href{https://doi.org/10.1088/1751-8113/46/28/285402}{\emph{J. Phys.}
  {\bfseries A46} (2013) 285402}
  [\href{https://arxiv.org/abs/1304.7985}{{\ttfamily 1304.7985}}].

\bibitem{Hubeny:2018ijt}
V.~E. Hubeny, M.~Rangamani and M.~Rota, \emph{{The holographic entropy
  arrangement}}, \href{https://doi.org/10.1002/prop.201900011}{\emph{Fortsch.
  Phys.} {\bfseries 67} (2019) 1900011}
  [\href{https://arxiv.org/abs/1812.08133}{{\ttfamily 1812.08133}}].

\bibitem{Hubeny:2018trv}
V.~E. Hubeny, M.~Rangamani and M.~Rota, \emph{{Holographic entropy relations}},
  \href{https://doi.org/10.1002/prop.201800067}{\emph{Fortsch. Phys.}
  {\bfseries 66} (2018) 1800067}
  [\href{https://arxiv.org/abs/1808.07871}{{\ttfamily 1808.07871}}].

\bibitem{Bao:2015bfa}
N.~Bao, S.~Nezami, H.~Ooguri, B.~Stoica, J.~Sully and M.~Walter, \emph{{The
  Holographic Entropy Cone}},
  \href{https://doi.org/10.1007/JHEP09(2015)130}{\emph{JHEP} {\bfseries 09}
  (2015) 130} [\href{https://arxiv.org/abs/1505.07839}{{\ttfamily
  1505.07839}}].

\bibitem{krtous}
P.~Krtou\v{s} and A.~Zelnikov, \emph{Minimal surfaces and entanglement entropy
  in anti-de {S}itter space}, {\emph{JHEP} {\bfseries 2014} (2014) }
  [\href{https://arxiv.org/abs/1406.7659}{{\ttfamily 1406.7659}}].

\bibitem{largeD}
R.~Emparan, R.~Suzuki and K.~Tanabe, \emph{The large {D} limit of general
  relativity}, {\emph{JHEP} {\bfseries 2013} (2013) }
  [\href{https://arxiv.org/abs/1302.6382}{{\ttfamily 1302.6382}}].

\bibitem{Headrick:2014cta}
M.~Headrick, V.~E. Hubeny, A.~Lawrence and M.~Rangamani, \emph{{Causality \&
  holographic entanglement entropy}},
  \href{https://doi.org/10.1007/JHEP12(2014)162}{\emph{JHEP} {\bfseries 12}
  (2014) 162} [\href{https://arxiv.org/abs/1408.6300}{{\ttfamily 1408.6300}}].

\bibitem{maximin}
A.~C. Wall, \emph{Maximin surfaces, and the strong subadditivity of the
  covariant holographic entanglement entropy}, {\emph{Class. Quantum Grav.}
  {\bfseries 31} (2014) 225007}
  [\href{https://arxiv.org/abs/1211.3494}{{\ttfamily 1211.3494}}].

\bibitem{Czech:2012bh}
B.~Czech, J.~L. Karczmarek, F.~Nogueira and M.~Van~Raamsdonk, \emph{The gravity
  dual of a density matrix},
  \href{https://doi.org/10.1088/0264-9381/29/15/155009}{\emph{Class. Quant.
  Grav.} {\bfseries 29} (2012) 155009}
  [\href{https://arxiv.org/abs/1204.1330}{{\ttfamily 1204.1330}}].

\bibitem{DHW}
X.~Dong, D.~Harlow and A.~C. Wall, \emph{Reconstruction of bulk operators
  within the entanglement wedge in gauge-gravity duality}, {\emph{Phys. Rev.
  Lett.} {\bfseries 117} (2016) 021601}
  [\href{https://arxiv.org/abs/1601.05416}{{\ttfamily 1601.05416}}].

\bibitem{noisyDHW}
J.~Cotler, P.~Hayden, G.~Penington, G.~Salton, B.~Swingle and M.~Walter,
  \emph{Entanglement wedge reconstruction via universal recovery channels},
  \href{https://arxiv.org/abs/1704.05839}{{\ttfamily 1704.05839}}.

\bibitem{Herzog:2014fra}
C.~P. Herzog, \emph{Universal thermal corrections to entanglement entropy for
  conformal field theories on spheres},
  \href{https://doi.org/10.1007/JHEP10(2014)028}{\emph{JHEP} {\bfseries 10}
  (2014) 028} [\href{https://arxiv.org/abs/1407.1358}{{\ttfamily 1407.1358}}].

\bibitem{mean-fields}
G.~A. Raggio and R.~F. Werner, \emph{Quantum statistical mechanics of general
  mean field systems}, {\emph{Helvetica Physica Acta} {\bfseries 62} (1989)
  980}.

\bibitem{mean-fields-1}
M.~R. Dowling, A.~C. Doherty and S.~D. Bartlett, \emph{Energy as an
  entanglement witness for quantum many-body systems}, {\emph{Phys. Rev. A}
  {\bfseries 70} (2004) 062113}
  [\href{https://arxiv.org/abs/quant-ph/0408086}{{\ttfamily
  quant-ph/0408086}}].

\bibitem{mean-fields-2}
A.~Osterloh and R.~Sch{\"u}tzhold, \emph{Monogamy of entanglement and improved
  mean-field ansatz for spin lattices}, {\emph{Phys. Rev. B} {\bfseries 91}
  (2015) 125114} [\href{https://arxiv.org/abs/1406.0311}{{\ttfamily
  1406.0311}}].

\bibitem{mean-fields-3}
C.~V. Kraus, M.~Lewenstein and J.~I. Cirac, \emph{Ground states of fermionic
  lattice hamiltonians with permutation symmetry}, {\emph{Phys. Rev. A}
  {\bfseries 88} (2013) 022335}
  [\href{https://arxiv.org/abs/1305.4577}{{\ttfamily 1305.4577}}].

\bibitem{qdF-1}
E.~St{\o}rmer, \emph{Symmetric states of infinite tensor products of
  $c^*$-algebras}, {\emph{J. Funct. Anal.} {\bfseries 3} (1969) 48}.

\bibitem{qdF-2}
R.~L. Hudson and G.~R. Moody, \emph{Locally normal symmetric states and an
  analogue of de {F}inetti's theorem}, {\emph{Probability Theory and Related
  Fields} {\bfseries 33} (1976) 343}.

\bibitem{qdF-3}
R.~K{\"o}nig and R.~Renner, \emph{A de {F}inetti representation for finite
  symmetric quantum states}, {\emph{J. Math. Phys.} {\bfseries 46} (2005)
  122108} [\href{https://arxiv.org/abs/quant-ph/0410229}{{\ttfamily
  quant-ph/0410229}}].

\bibitem{big-qdF}
M.~Christandl, R.~K{\"o}nig, G.~Mitchison and R.~Renner, \emph{One-and-a-half
  quantum de {F}inetti theorems}, {\emph{Comm. Math. Phys.} {\bfseries 273}
  (2007) 473} [\href{https://arxiv.org/abs/quant-ph/0602130}{{\ttfamily
  quant-ph/0602130}}].

\bibitem{LOCC-decoupling}
F.~G. Brandao, M.~Christandl and J.~Yard, \emph{A quasipolynomial-time
  algorithm for the quantum separability problem}, {\emph{Proceedings of ACM
  Symposium on Theory of Computation} (2011) 343}
  [\href{https://arxiv.org/abs/1011.2751}{{\ttfamily 1011.2751}}].

\bibitem{MMI}
P.~Hayden, M.~Headrick and A.~Maloney, \emph{Holographic mutual information is
  monogamous}, {\emph{Phys. Rev. D} {\bfseries 87} (2013) 046003}
  [\href{https://arxiv.org/abs/1107.2940}{{\ttfamily 1107.2940}}].

\bibitem{Faulkner:2013ica}
T.~Faulkner, M.~Guica, T.~Hartman, R.~C. Myers and M.~Van~Raamsdonk,
  \emph{{Gravitation from Entanglement in Holographic CFTs}},
  \href{https://doi.org/10.1007/JHEP03(2014)051}{\emph{JHEP} {\bfseries 03}
  (2014) 051} [\href{https://arxiv.org/abs/1312.7856}{{\ttfamily 1312.7856}}].

\bibitem{other-spacetimes-1}
E.~Tonni, \emph{Holographic entanglement entropy: near horizon geometry and
  disconnected regions}, {\emph{JHEP} {\bfseries 2011} (2011) }
  [\href{https://arxiv.org/abs/1011.0166}{{\ttfamily 1011.0166}}].

\bibitem{other-spacetimes-2}
W.~Fischler, A.~Kundu and S.~Kundu, \emph{Holographic mutual information at
  finite temperature}, {\emph{Phys. Rev. D} {\bfseries 87} (2013) 126012}
  [\href{https://arxiv.org/abs/1212.4764}{{\ttfamily 1212.4764}}].

\bibitem{other-spacetimes-3}
S.~Kundu and J.~F. Pedraza, \emph{Aspects of holographic entanglement at finite
  temperature and chemical potential}, {\emph{JHEP} {\bfseries 2016} (2016) }
  [\href{https://arxiv.org/abs/1602.07353}{{\ttfamily 1602.07353}}].

\bibitem{other-spacetimes-4}
M.~Ghodrati, X.-M. Kuang, B.~Wang, C.-Y. Zhang and Y.-T. Zhou, \emph{Aspects of
  holographic entanglement at finite temperature and chemical potential},
  {\emph{JHEP} {\bfseries 2019} (2019) }
  [\href{https://arxiv.org/abs/1902.02475}{{\ttfamily 1902.02475}}].

\bibitem{Shenker:2013pqa}
S.~H. Shenker and D.~Stanford, \emph{{Black holes and the butterfly effect}},
  \href{https://doi.org/10.1007/JHEP03(2014)067}{\emph{JHEP} {\bfseries 03}
  (2014) 067} [\href{https://arxiv.org/abs/1306.0622}{{\ttfamily 1306.0622}}].

\bibitem{large-d-blocks}
A.~L. Fitzpatrick, J.~Kaplan and D.~Polanda, \emph{Conformal blocks in the
  large {D} limit}, {\emph{JHEP} {\bfseries 2013} (2013) }
  [\href{https://arxiv.org/abs/1305.0004}{{\ttfamily 1305.0004}}].

\bibitem{Strominger:1981jg}
A.~Strominger, \emph{{The Inverse Dimensional Expansion in Quantum Gravity}},
  \href{https://doi.org/10.1103/PhysRevD.24.3082}{\emph{Phys. Rev.} {\bfseries
  D24} (1981) 3082}.

\bibitem{CHRR}
S.~{Hernandez Cuenca}, V.~E. Hubeny, M.~Rangamani and M.~Rota, \emph{The
  quantum marginal independence problem}, {\emph{(forthcoming)} }.

\bibitem{Swingle:2014uza}
B.~Swingle and M.~Van~Raamsdonk, \emph{{Universality of Gravity from
  Entanglement}},  \href{https://arxiv.org/abs/1405.2933}{{\ttfamily
  1405.2933}}.

\end{thebibliography}\endgroup

\end{document}